\documentclass{aa}
\usepackage[varg]{txfonts}
\usepackage{natbib}
\usepackage{subcaption}
\usepackage[colorlinks=true,citecolor=blue]{hyperref}%
\usepackage{graphicx}

\bibpunct{(}{)}{;}{a}{}{,} 

\newcommand{\Planck}{{\em Planck}}
\newcommand{\bh}{b_{\rm h}}
\newcommand{\Pe}{P_\textrm{e}}
\newcommand{\rs}{r_\textrm{s}}
\newcommand{\bg}{b_\textrm{g}}
\newcommand{\kB}{k_\textrm{B}}
\newcommand{\sigmaT}{\sigma_\textrm{T}}
\newcommand{\me}{m_\textrm{e}}

\begin{document}

\title{Cosmic census: Relative distributions of dark matter, galaxies, and diffuse gas}

\author{Raphaël Kou\inst{1}
  \and James G. Bartlett\inst{1}} 

\institute{Université Paris Cité, CNRS, Astroparticule et Cosmologie, F-75013 Paris, France} 

\date{Received / Accepted}

\abstract{Galaxies, diffuse gas, and dark matter make up the cosmic web that defines the large-scale structure of the Universe. 
We constrained the joint distribution of these constituents by cross-correlating galaxy samples binned by stellar mass from the Sloan Digital Sky Survey CMASS catalog with maps of lensing convergence and the thermal Sunyaev-Zeldovich (tSZ) effect from the \Planck\ mission.  Fitting a halo-based model to our measured angular power spectra (galaxy-galaxy, galaxy-lensing convergence, and galaxy-tSZ) at a median redshift of $z=0.53$, we detected variation with stellar mass of the galaxy satellite fraction and galaxy spatial distribution within host halos. We find a tSZ-halo hydrostatic mass bias, $\bh$, such that  $(1-\bh)=0.6\pm0.05$, with a hint of a larger bias, $\bh$, at the high stellar mass end. The normalization of the galaxy-cosmic microwave background lensing convergence cross-power spectrum shows that galaxies trace the matter distribution without an indication of stochasticity ($A=0.98\pm 0.09$). We forecast that next-generation cosmic microwave background experiments will improve the constraints on the hydrostatic bias by a factor of two and will be able to constrain the small-scale distribution of dark matter, hence informing the theory of feedback processes.}

\maketitle

\section{Introduction}
The standard cosmological model successfully explains current observations with a spatially flat universe composed of 68\% dark energy and 32\% matter, the latter dominated by nonbaryonic cold dark matter (80\%) and the rest as ordinary baryonic matter \citep[e.g.,][]{2020A&A...641A...6P}.  Gravity  arranges dark matter into filaments 
that the baryons, in the form of diffuse gas and galaxies, fill out, the whole making the cosmic web \citep{CosmicWeb}. Nongravitational mechanisms involving cooling and feedback drive the galaxy evolution and determine the relative distributions of gas, galaxies, and dark matter within the web.  Measurements of their joint distribution tell us about these mechanisms. Notably, only about 10-15\% of the baryons reside as stars and gas in galaxies today, implying that  feedback prevented the majority from cooling into stars and dispersed it as a diffuse gas into  galaxy halos (the circumgalactic medium, CGM) and the broadly distributed intergalactic medium \citep[IGM;][]{FukugitaPeebles2006, Tumlinson+2017}. Understanding these processes is central to galaxy formation theory.  


Large astronomical surveys spanning the electromagnetic spectrum now enable a cosmic census of the dark matter, galaxy, and gas distributions. Spectroscopic galaxy surveys provide detailed information on the galaxy distribution as a function of galaxy properties, such as stellar mass. Over large scales, galaxies trace the dark matter field with a linear bias \citep[for a review, see][]{2018PhR...733....1D}. This bias is crucial in galaxy clustering studies since it is degenerate with $\sigma_8$, the amplitude of the matter fluctuations. On small scales, the relation between galaxies and the underlying dark matter distribution becomes more complex and nonlinear.  

Gravitational lensing directly probes the dark matter distribution. Lensing distorts the images of distant galaxies and permits reconstruction of the foreground mass distribution \citep[e.g.,][]{1993ApJ...404..441K,2001PhR...340..291B,2018MNRAS.475.3165C}. Lensing also leaves a characteristic imprint on the anisotropies of the cosmic microwave background (CMB), which presents a back-light probing the matter distribution across the observable Universe \citep{2006PhR...429....1L}. The effect was first detected through a cross-correlation of Wilkinson Microwave Anisotropy Probe (WMAP) sky maps and tracers of large-scale structure  \citep{2007PhRvD..76d3510S, 2008PhRvD..78d3520H}. The most recent work from \Planck\ \citep{2020A&A...641A...8P} measured the lensing signal in the CMB at $40\sigma$ significance. The South Pole Telescope (SPT) detected the lensing amplitude at $39\sigma$ significance using  temperature and polarization measurements, and at $10.1\sigma$ significance when using only polarization; and the Atacama Cosmology Telescope (ACT) has constructed a lensing map covering 2100 square degrees \citep{2019ApJ...884...70W,ACTLensing2014,2021MNRAS.500.2250D}. 

The distribution of diffuse ionized gas can be probed through the thermal and kinetic Sunyaev-Zeldovich (tSZ and kSZ) effects. Energy transferred to CMB photons when scattered by free hot electrons creates the unique spectral signature of the tSZ effect relative to the pure thermal CMB spectrum. The amplitude of the tSZ measures the gas pressure along the line of sight.
Any bulk peculiar motion of the gas relative to the comoving frame modifies the CMB brightness via the Doppler effect to generate the kSZ signal, which  is proportional to the product of the bulk velocity and the Thomson optical depth along the line of sight \citep{Carlstrom+2002, 1972CoASP...4..173S}. 

With galaxy and CMB surveys now covering large overlapping areas of extragalactic sky, these three probes can be combined to study the joint distributions of dark matter, galaxies, and gas. Cross-correlations between  CMB lensing and galaxies have  been performed in a number of studies \citep[e.g.,][]{2015A&A...584A..53K,2015ApJ...802...64B,2016MNRAS.460.4098P,2017MNRAS.464.2120S,2018MNRAS.480.5386D,2019ApJ...886...38A,2021MNRAS.501.1481H,2021arXiv210602551S,2021MNRAS.501.6181K,2021MNRAS.500.2250D}. Most studies aim either to measure the cross-correlation amplitude (which should be equal to $1$ in the standard cosmological model) or to constrain cosmological parameters (primarily $\sigma_8$ and $\Omega_m$). Galaxies have also been cross-correlated with the tSZ effect to perform a (sometimes tomographic) measurement of the gas pressure \citep{2017MNRAS.467.2315V,2020MNRAS.491.5464K}, or to measure the mass bias \citep{2018MNRAS.480.3928M}. \cite{PIPXI}, and \cite{Greco2015} stacked \Planck\ tSZ measurements of the locally brightest galaxies to measure the  relation between the galaxy mass and the thermal energy of ionized gas in and around galaxy halos. \citet{2021PhRvD.103f3514A} probed the gas thermodynamics of CMASS galaxies using cross-correlations with the measurements of tSZ and kSZ effects from the ACT, whereas \citet{2021PhRvD.104d3503V} also used these effects to probe the baryon content of SDSS DR15 galaxies. 
Other studies have cross-correlated the tSZ effect with CMB lensing \citep{2014JCAP...02..030H,2022PhRvD.105l3526P}, and \citet{2021A&A...651A..76Y} cross-correlated the three probes to constrain the redshift evolution of the galaxy and gas pressure bias factors.

Most cross-correlation studies aim at constraining cosmological parameters or at testing fundamental physics; for instance,  cross-correlations break the degeneracy between the galaxy bias and cosmology. As mentioned, some studies binned the galaxies in redshift, for example, to obtain tomographic information on dark matter or gas pressure distributions. A number of studies have examined the clustering of galaxies binned according to their properties (color, stellar mass, and star formation rate) \citep{2008MNRAS.385.1635S,2013A&A...557A..17M,2018A&A...612A..42D,2019ApJ...871..147G,2020ApJ...904..128I,2021A&A...646A..71S}. By binning on galaxy properties, it is possible not only to study the galaxy distribution, but also to probe how the dark matter and gas (or any other cross-correlated field) distributions depend on galaxy properties.

In this work, we perform a cosmic census of the joint distribution of different matter components with the aim of better understanding their relations. We study the dependence of galaxy, gas, and mass distributions on stellar mass, using a halo model formalism to interpret our measurements of galaxy-galaxy, galaxy-convergence, and galaxy-tSZ cross-spectra. The halo model formalism builds upon~\citet{2002PhR...372....1C}, \citet{2013MNRAS.430..767C}, \citet{2013MNRAS.430..725V} and on \citet{2015ApJ...806....2M}, who also performed an analysis of the clustering of   CMASS galaxies binned by stellar mass to constrain the cosmology.

The paper is organized as follows: Sect.~\ref{sec:data} details the data sets we used, and Sect.~\ref{sec:measurements} presents our method for measuring the cross-power spectra. Section~\ref{sec:modeling} describes our   modeling approach. We present and discuss our results in Sect.~\ref{sec:results} and conclude in Sect.~\ref{sec:conclusion}.
The cosmology used in this paper is fixed and follows \citet{2020A&A...641A...6P} (TT, TE, EE+lowE+lensing+BAO, where $T$ and $E$ denote the CMB temperature and polarization, and BAO stands for baryon acoustic oscillation).

\section{Data}
\label{sec:data}
\subsection{Galaxy catalog}
We used the Sloan Digital Sky Survey (SDSS) Baryon Oscillation Spectroscopic Survey (BOSS) Data Release 12 (DR12) CMASS galaxy sample\footnote{https://data.sdss.org/sas/dr12/boss/} \citep{2015ApJS..219...12A}. The BOSS survey observed galaxies up to redshift $z=0.7$ and was divided into two samples (LOWZ and CMASS). The LOWZ sample contains low-redshift galaxies, and the CMASS sample consists of galaxies in the redshift range $0.43 < z < 0.7$.

\subsubsection{Subsample definition}
\begin{figure}
  \resizebox{\hsize}{!}{\includegraphics{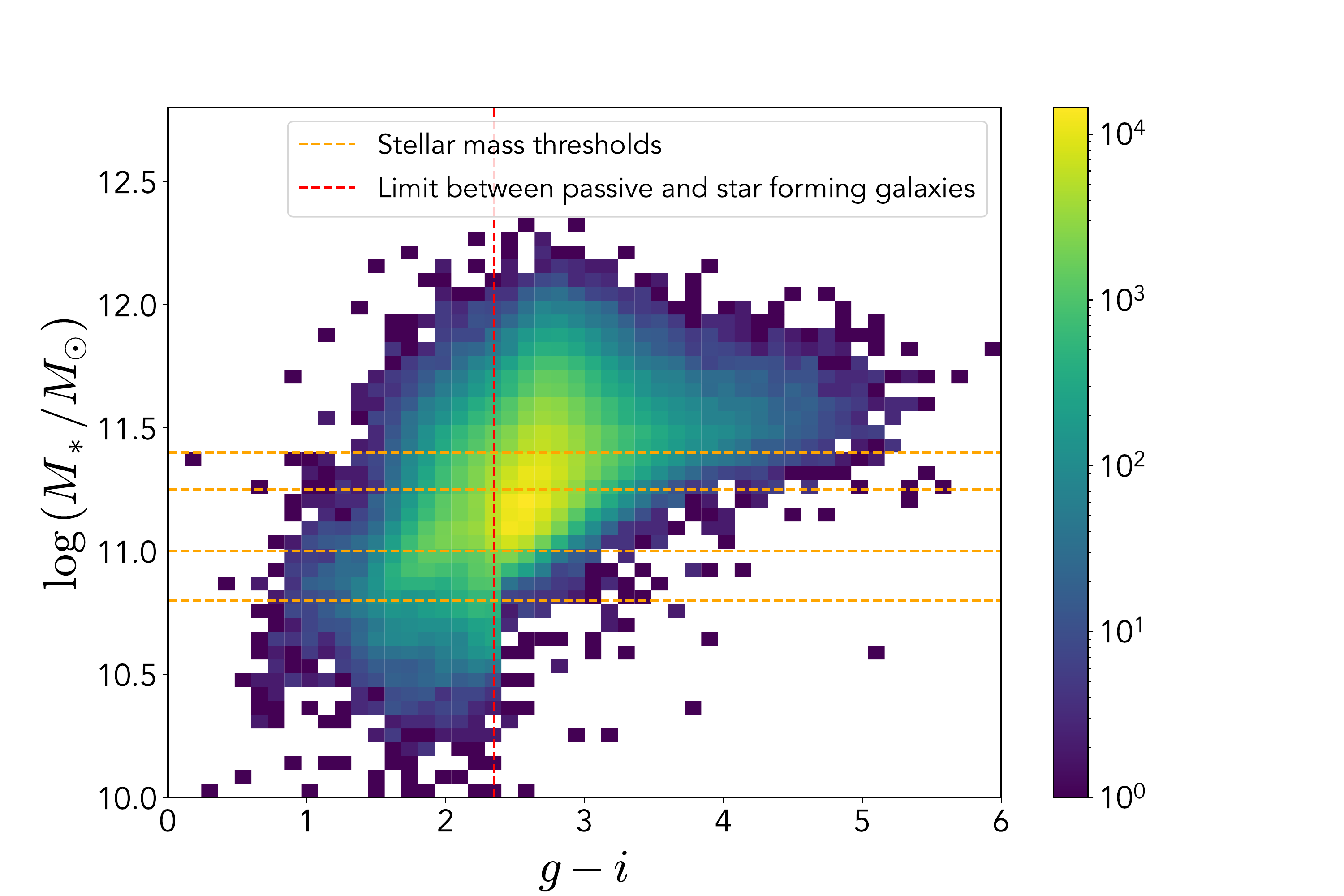}}
  \caption{Number of galaxies in a color ($g-i$) and stellar mass pixel. The horizontal orange lines show the stellar mass thresholds we used for the bins. The vertical red line indicates the color criteria we used to determine whether a galaxy should be considered as passive or star forming.}\label{fig:distribution_galaxy_color_stellar_mass}
\end{figure}

 We  restricted our work to galaxies in the redshift range $0.47<z<0.59$, following \citet{2015ApJ...806....2M}, in order to work with a sample of nearly constant (as a function of redshift) number density. 
We also followed \citet{2015ApJ...806....2M} and used the Portsmouth stellar population synthesis code \citep{2013MNRAS.435.2764M} to retrieve galaxy stellar masses. \citet{2013MNRAS.435.2764M} obtained the stellar masses by fitting model spectral energy distributions to u, g, r, i, z magnitudes. They developed two different models, depending on whether a galaxy was assumed to be star forming or passive. We used their results for a Kroupa \citep{2001MNRAS.322..231K} initial mass function and chose between the star-forming or passive catalogs according to the simple color criteria they advocate: a star-forming catalog for galaxies with $g-i<2.35,$ and the passive model otherwise. We then defined four subsamples by stellar mass, $M_*$, such that $\log{M_*/M_\odot}>10.8$, $\log{M_*/M_\odot}>11.1$, $\log{M_*/M_\odot}>11.25,$ and $\log{M_*/M_\odot}>11.4$. The corresponding number of galaxies in each subsample is $473596$, $396298$, $250964,$ and $124493$, respectively. Figure~\ref{fig:distribution_galaxy_color_stellar_mass} shows the distribution of CMASS galaxies (in the restricted reshift range $0.47<z<0.59$) as a function of their color $g-i$ and the stellar mass given by \citet{2013MNRAS.435.2764M}. Most galaxies are passive, especially at high stellar mass. In the first (second, third, and fourth) stellar mass bin, only $19.51\%$ ($13.77\%$, $11.27\%,$ and $8.90\%$)  of the galaxies are star forming.

\subsubsection{Galaxy maps and masks}
We constructed maps and corresponding masks of the galaxy distribution from the catalogs following an approach similar to that of  \citet{2016MNRAS.455.1553R} and \citet{2019MNRAS.485..326L}. The SDSS collaboration released acceptance and veto masks in {\sl MANGLE}\footnote{https://space.mit.edu/~molly/mangle/} format \citep{2008MNRAS.387.1391S}. The acceptance mask represents the completeness of the observations in a given region of the sky; it represents the fraction of galaxies for which a spectrum could be obtained compared to the total number of targets. The veto masks exclude regions of the sky in which no galaxies were observed, for example, because of point spread function (PSF) modeling failures, reduction pipeline timeout due to too many blended objects, regions around the center posts of the plates, or where higher priority objects were observed. We refer to \cite{2014MNRAS.441...24A} for details. In total, about $5\%$ of the covered sky is excluded by the veto masks.

We transformed these {\sl MANGLE} masks into a single high-resolution binary mask in {\sl HEALPix}\footnote{http://healpix.sf.net} \citep{healpix, Zonca2019} format using its Python implementation {\sl healpy}  with $\textrm{N}_\textrm{SIDE}=8192$ (corresponding to a pixel size of $0.43$ arcmin), such that any pixel from a veto mask or with a completeness (in the acceptance mask) lower than $0.7$ was rejected. We then degraded this binary mask to a lower-resolution mask ($\textrm{N}_\textrm{SIDE}=4096$, corresponding to a pixel size of $0.86$ arcmin) where the value of each pixel is the mean of its subpixels in the high-resolution binary mask. This yielded our final completeness mask, whose values are noted $C_{\textrm{pix}}$ following \citet{2019MNRAS.485..326L}. 

As in their study (even though the resolution of the mask we used is different), when we computed the galaxy maps, we only accepted pixels with completeness $C_{\textrm{pix}}>0.8$. This condition is different from the previous cut that was performed on completenesses lower than $0.7$ because that cut was made before to degrade the mask resolution to $\textrm{N}_\textrm{SIDE}=4096$. As a result, some regions that were not excluded from the high-resolution mask can eventually be rejected in the lower-resolution mask if the neighboring regions are not complete enough. Figure~\ref{fig:masks} shows the mask.

Each galaxy was weighted to account for a number of observational effects, such as fibre collisions or redshift failures. Following \cite{2014MNRAS.441...24A}, we assigned a weight to each galaxy,
\begin{align}
    w_{\textrm{tot}} = (w_{\textrm{cp}}+w_{\textrm{zf}}-1)w_{\textrm{star}}w_{\textrm{see},}
\end{align}
where $w_\textrm{cp}$ and $w_\textrm{zf}$ account for missing redshifts due to close pairs and redshift failures by upweighting the nearest neighbor galaxy, respectively. The $w_\textrm{star}$ and $w_\textrm{see}$ weights correct for the fact that fewer galaxies are observed in regions with high stellar densities and due to  seeing conditions, respectively \citep[for more details, see][]{2012MNRAS.424..564R}. We then constructed a galaxy count map as
\begin{align}
    n_p = \left\{
    \begin{array}{ll}
        \frac{1}{C_\textrm{pix}^p}\sum_{i\in p} w_\textrm{tot}^i & \mbox{if } C_\textrm{pix}^p > 0.8 \\
        0 & \mbox{otherwise},
    \end{array}
\right.
\end{align}
where $n_p$ is the weighted number of galaxies in pixel $p$. As mentioned previously, we dropped pixels with a completeness lower than $0.8$. Finally, we built overdensity maps defined as
\begin{align}\label{eq:overdensity_equation}
    \delta_p = \left(\frac{n_p}{\bar{n}}-1\right),
\end{align}
where $\bar{n}$ is the mean of the weighted galaxy count map.

\subsection{(CMB lensing) Convergence map}
We used the CMB lensing map provided by the Planck\footnote{http://pla.esac.esa.int/pla/} Collaboration in their 2018 release and described in \citet{2020A&A...641A...8P}. The map was computed using a minimum-variance estimate from temperature and polarization data and covers $67\%$ of the sky. It provides spherical harmonic coefficients for the lensing convergence, $\kappa$, with $\textrm{N}_\textrm{SIDE}=4096$, together with the corresponding mask given at $\textrm{N}_\textrm{SIDE}=2048$. The convergence is related to the lensing potential, $\phi$, through
\begin{align}
    \kappa_{\ell m} = \frac{\ell(\ell+1)}{2}\phi_{\ell m}.
\end{align}
We used {\sl healpy} to transform these spherical harmonics into a real-space sky map.

\subsection{tSZ map}
For the tSZ effect, we used the all-sky Compton-$y$ parameter map given in \citet{2016A&A...594A..22P} and released as part of the 2015 \Planck\ data release. The map and associated mask are provided in {\sl HEALPix} format with $\textrm{N}_\textrm{SIDE}=2048$. We applied the \Planck\ $60$\% Galactic mask combined with the point-source mask to remove Galactic emission and radio and infrared sources contaminating the $y$ map. We also corrected for cosmic infrared background (CIB) contamination as described in Sect.~\ref{sec:CIB}.

\begin{figure*}
    \centering
    \begin{subfigure}[b]{5.66cm}
        \includegraphics[width=\textwidth]{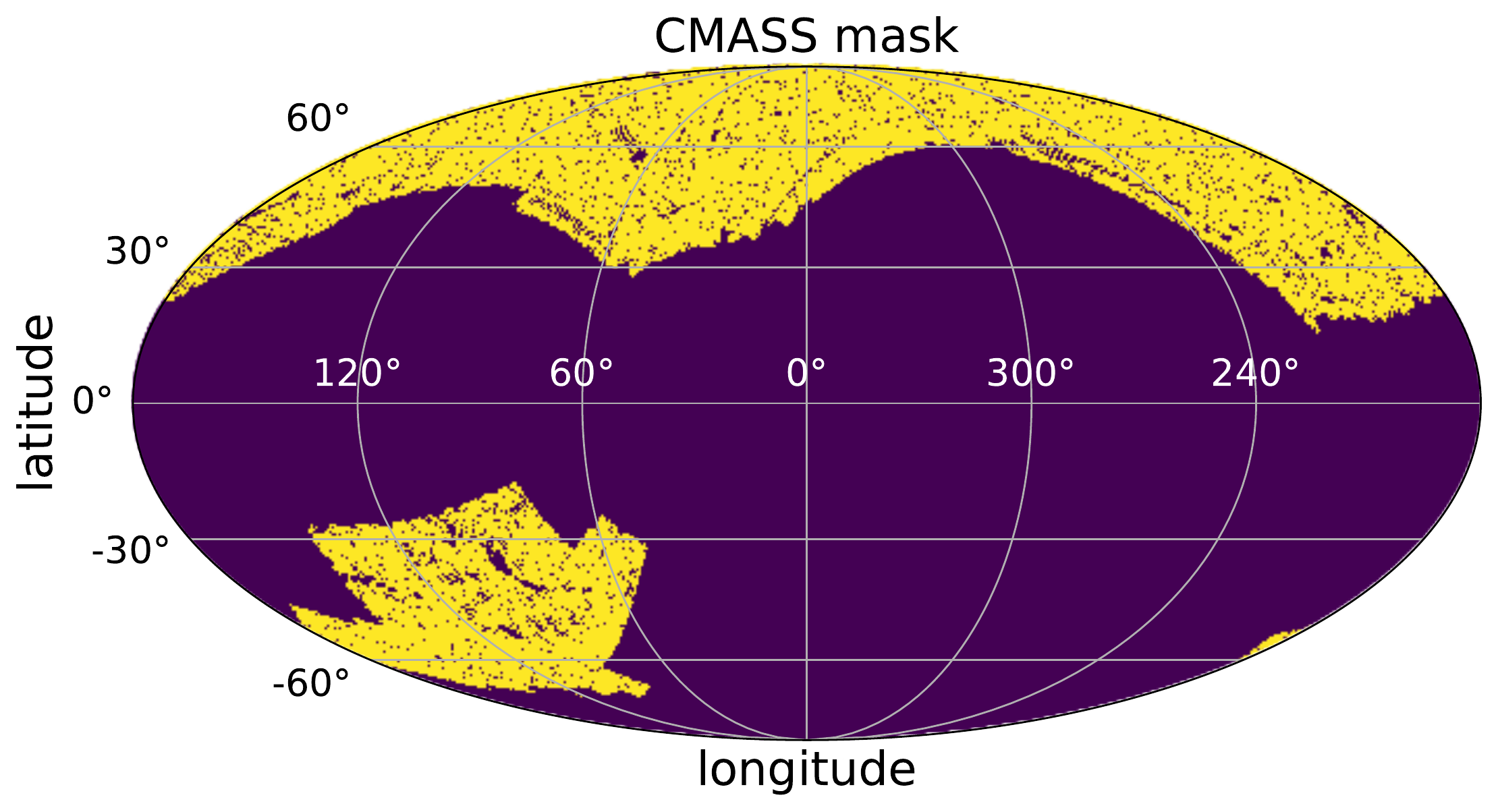}
    \end{subfigure}
    \begin{subfigure}[b]{5.66cm}
        \includegraphics[width=\textwidth]{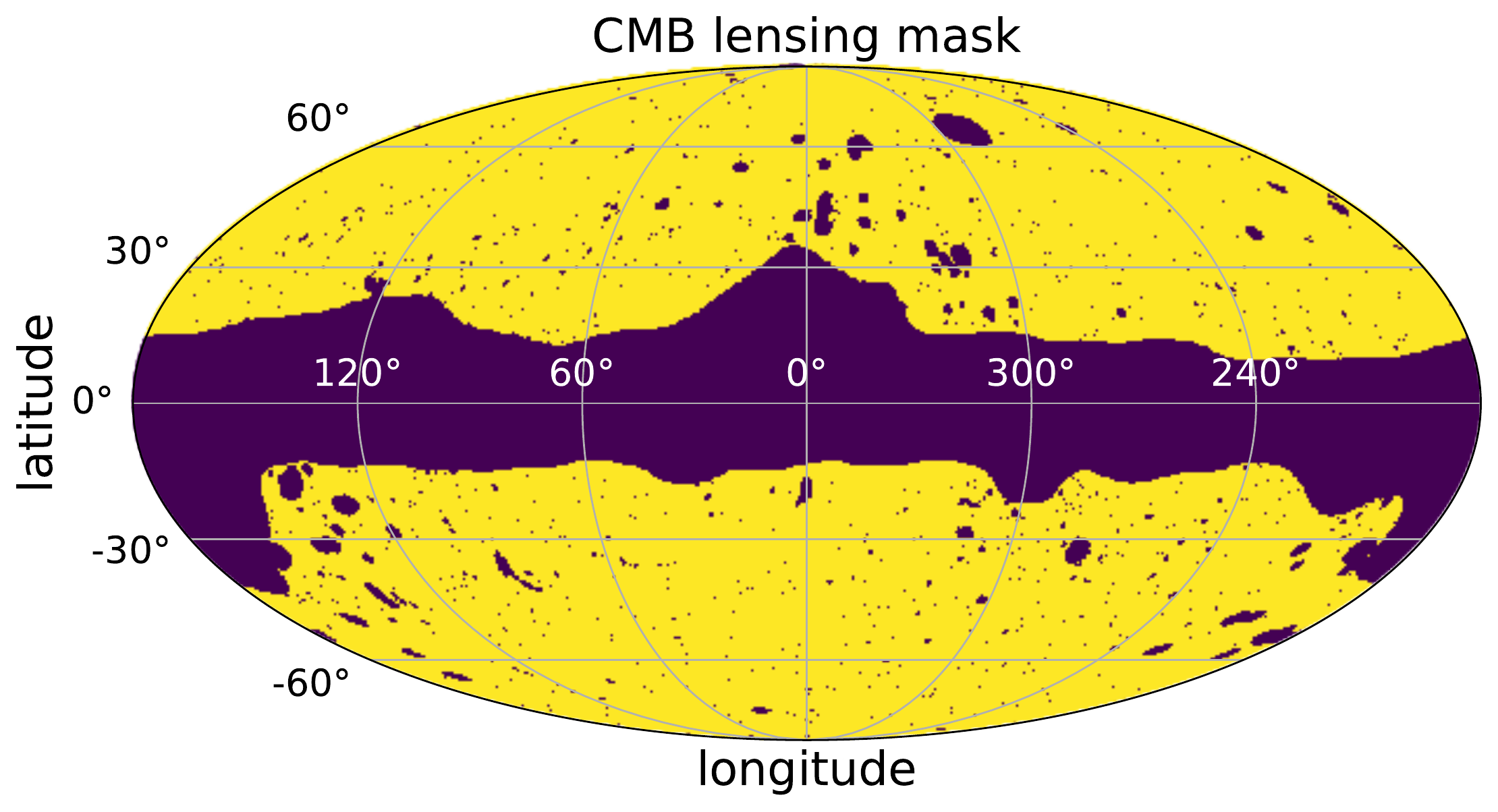}
    \end{subfigure}
    \begin{subfigure}[b]{5.66cm}
        \includegraphics[width=\textwidth]{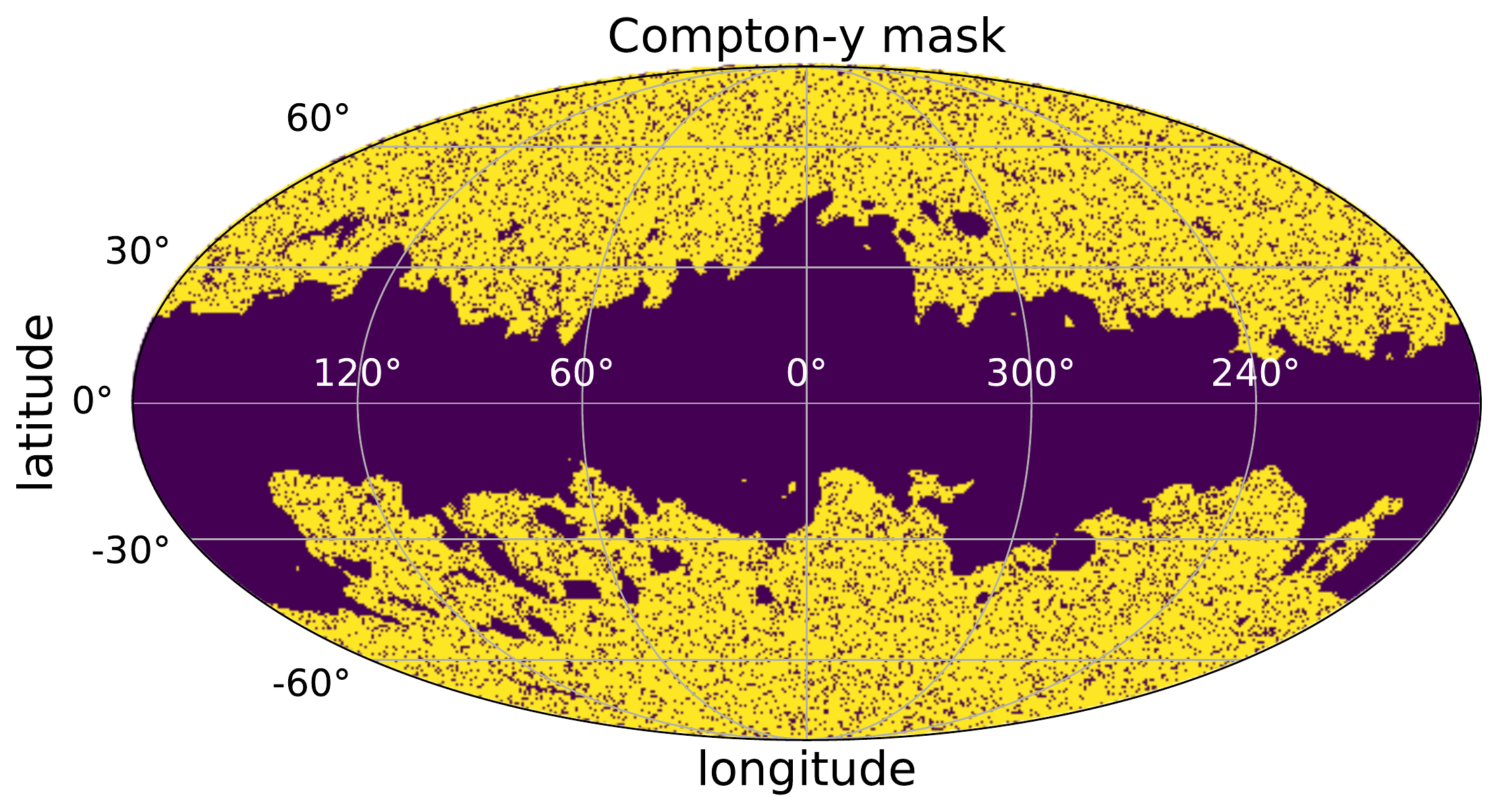}
    \end{subfigure}
    \caption{CMASS, CMB lensing, and Compton-$y$ masks in Galactic coordinates using the Mollweide projection.}.
    \label{fig:masks}
\end{figure*}

\section{Measurement method}
\label{sec:measurements}
\subsection{Estimating the power spectra}
We decomposed each map, $a$ (galaxy density map, convergence map, and Compton-$y$ map), a scalar function of the sky position $\hat{\theta}$, into spherical harmonics as
\begin{align}
    a(\hat{\theta}) = \sum_{\ell m} a_{\ell m}Y_{\ell m}(\hat{\theta}).
\end{align}
The angular power (cross)-spectrum between two maps ($a$ and $a'$), $C_\ell^{aa'}$, is defined by
\begin{align}
    \langle a_{\ell m}a_{\ell'm'}^{'*} \rangle = C_\ell^{aa'}\delta_{\ell\ell'}\delta_{mm'},
\end{align}
where angle brackets indicate an average over the theoretical ensemble of possible skies, and $\delta$ represents the Kronecker delta function.  An unbiased estimator of the power spectrum, given full-sky coverage, is
\begin{align}
    C_\ell^{aa'} = \frac{1}{2\ell+1}\sum_{m=-\ell}^\ell a_{\ell m}a_{\ell m}^{'*}.
\end{align}
Incomplete sky coverage and masking, however, limit our maps to less than full-sky coverage, as shown in Fig.~\ref{fig:masks}, coupling the different multipoles and reducing the resolution in harmonic space. We therefore binned the multipoles and calculated the bin-bin coupling matrix using the {\sl NaMaster}\footnote{https://github.com/LSSTDESC/NaMaster} code from \citet{2019MNRAS.484.4127A}. 

The code also corrects for beam smoothing of the Compton-$y$ map. Following \citet{2016A&A...594A..22P}, we adopted a circular Gaussian beam of $10$\,arcmin full width at half maximum (FWHM), resulting in beam harmonic coefficients
\begin{align}
    b_l = \exp{-\ell(\ell+1)\sigma^2/2},
\end{align}
with $\sigma = \textrm{FWHM}/\sqrt{8\ln{2}}$.
Additionally, we used the {\sl NaMaster} apodization function with a smoothing scale of $10$\,arcmin to apodize the tSZ and lensing convergence masks. 

The multipole binning followed a linear scheme from $\ell=30$ to $190$ and logarithmic from $190$ to $3800$. We used the same scheme for every power spectrum, except for the cross-spectrum between galaxies and Compton-$y$ parameter, for which we only considered multipoles below $\ell = 1250$ because, even though we corrected for beam smoothing, the Gaussian approximation might bias the high multipoles since the beam smoothing mainly affects the small scales (i.e., high multipoles). Moreover, there may remain some contamination from point sources and CIB (see Sect.~\ref{sec:CIB}).

\subsection{Computing the noise bias }
The only noise spectrum we computed is the galaxy shot noise because we only considered auto-power spectra for galaxies, where the shot noise contributes a noise bias term. The noise being uncorrelated between the different maps, however, means that there is no noise bias term in the cross-power spectra between the galaxy map and the maps of CMB lensing and the tSZ effect. The galaxy shot noise is  given by 
\begin{align}
    N_l^{gg} = \frac{4\pi f_{\textrm{sky}}}{N}, 
\end{align}
where $N$ is the weighted number of galaxies in each stellar mass bin, and $f_{\textrm{sky}}$ is the observed sky fraction.

\subsection{Covariance estimate}\label{sec:covariance}
In order to estimate the covariance of our measurements, we used the formula derived by \citet{2021arXiv210602551S}, which is a generalization of the usual Gaussian covariance matrix that allows us to use the full sky area observed by each probe instead of restricting the three probes to their common overlapping sky region.  The expression for the covariance matrix is given by
\begin{multline}\label{eq:covariance}
    {\rm Cov}_{LL'}^{AB,CD} = \frac{\delta_{LL'}}{(2\ell_L+1)\Delta \ell f_\textrm{sky}^{AB}f_\textrm{sky}^{CD}}\biggl[ f_\textrm{sky}^{AC,BD}\left(C_L^{AC}+N_L^{AC}\right) \\
    \times\left(C_L^{BD}+N_L^{BD}\right)+ f_\textrm{sky}^{AD,BC}\left(C_L^{AD}+N_L^{AD}\right)\left(C_L^{BC}+N_L^{BC}\right)\biggr],
\end{multline}
where $A$,$         B$, $C$, and $D$ are indices for galaxies, $g$, CMB convergence, $\kappa$, or Compton-$y$ parameter, $y$; $\Delta \ell$ is the multipole bin-width; and $f_\textrm{sky}^{AB}$ denotes the common sky fraction covered by fields $A$ and $B$.

\subsection{CIB decontamination}\label{sec:CIB}
 Redshifted thermal dust emission from star-forming galaxies generates the  cosmic infrared background (CIB) that  contaminates measurements of the tSZ signal. Since the CIB comes from galaxies, it traces large-scale structures and likely correlates with our other observations.  We used a method proposed in a number of studies \citep[e.g.,][]{2014JCAP...02..030H,2017MNRAS.467.2315V,2018PhRvD..97f3514A,2020MNRAS.491.5464K,2021A&A...651A..76Y} to mitigate this CIB contamination. 
 
 The observed $y$ map was modeled as
\begin{align}
    y_\textrm{obs}(\hat{\theta}) = y_\textrm{true}(\hat{\theta})+\epsilon_\textrm{CIB}c(\hat{\theta}),
\end{align}
where $y_\textrm{obs}(\hat{\theta})$ and $y_\textrm{true}(\hat{\theta})$ are the observed and decontaminated $y$ maps, respectively, and $c(\hat{\theta})$ is the CIB contamination weighted by some coefficient $\epsilon_\textrm{CIB}$. The cross-power spectrum then becomes
\begin{align}
    C_\ell^{gy,\textrm{obs}} = C_\ell^{gy,\textrm{true}}+\epsilon_\textrm{CIB}C_\ell^{gc}.
\end{align}
To estimate $C_\ell^{gc}$ , we cross-correlated our galaxy maps with the \Planck\ $545$ GHz map that was used as a template for the CIB contamination. \citet{2018PhRvD..97f3514A} then estimated the cross-correlation between this map and the Compton-$y$ map such that 
\begin{align}
    C_\ell^{cy,\textrm{obs}} = C_\ell^{cy,\textrm{true}}+\epsilon_\textrm{CIB}C_\ell^{cc}.
\end{align} Using models \citep{2014A&A...571A..30P,2016A&A...594A..23P} for $C_\ell^{cy,\textrm{true}}$ and $C_\ell^{cc}$, \citet{2018PhRvD..97f3514A} determined $\epsilon_\textrm{CIB}=(2.3 \pm 6.6) \times 10^{-7} (\textrm{MJy/sr})^{-1}$. 

\begin{figure}
   \resizebox{\hsize}{!}{\includegraphics{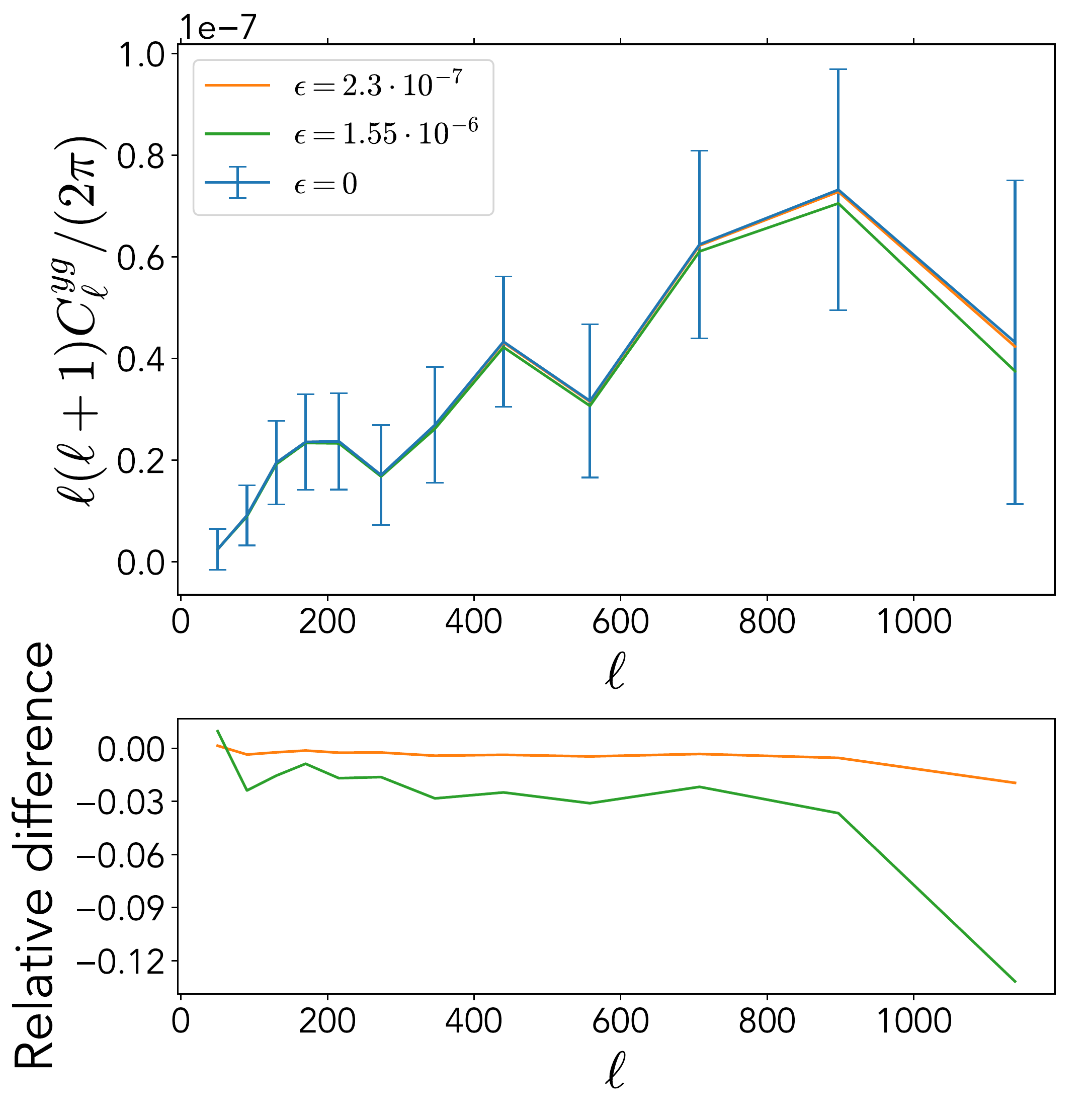}}
   \caption{Impact of CIB decontamination on $yg$ cross-power spectrum. \textit{Top}: Measured $yg$ cross-power spectrum (in blue) along with the associated uncertainties. The orange (green) curve is the CIB-decontaminated curve using the best-fit (the  2$\sigma$ upper) value for $\epsilon_\textrm{CIB}$.
   \textit{Bottom}: Relative difference between the CIB-decontaminated curves and our raw measured $yg$ cross-power spectrum. On the scales we considered, the effect of the CIB decontamination is $2.0\%$ at most with the best-fit value of $\epsilon_\textrm{CIB}$, and $13.2\%$ at most with the 2$\sigma$ upper value.}\label{fig:cl_yg_epsilon}
\end{figure}

We used this value to remove the CIB bias term. For our galaxy sample, this resulted in only a small correction ($2\%$ at most) at high multipoles. Given the high uncertainty in  $\epsilon_\textrm{CIB}$, we also subtracted the bias term using the 2$\sigma$ upper value of $\epsilon_\textrm{CIB}$, and this again remained a small correction (up to $13.2\%$) compared to the uncertainties on our measurements (see Fig.~\ref{fig:cl_yg_epsilon}). At higher multipoles or with the increasing accuracy of future CMB experiments, the CIB contamination could become an issue for this type of study. 

\section{Modeling approach}
\label{sec:modeling}
\subsection{Power spectrum line-of-sight projection}
To model our angular power spectra, we computed the three-dimensional theoretical power spectra (see Sect.~\ref{sec:halo_model}) for each probe and projected them along the line of sight using the Limber approximation \citep{1953ApJ...117..134L}, which is applicable except for the lowest multipoles, with kernels $W^A(z)$ adapted to each probe, 
\begin{align}\label{eq:los_projection}
    C_\ell^{AB} = \int \frac{dz}{c}\frac{H(z)}{\chi(z)^2}W_A(z)W_B(z)P^{AB}(k=\frac{\ell}{\chi(z)},z),
\end{align}
where $P^{AB}$ is the three-dimensional cross-power spectrum between tracers $A$ and $B$. In this expression, $\chi(z)$ is the comoving distance to redshift $z$, $H(z)$ is the Hubble parameter at $z$, $c$ is the speed of light, and $k$ is the comoving wavenumber. 

The kernels for the different probes (galaxies, CMB lensing convergence, and Compton-$y$ parameter) are given by
\begin{align}
    W_g(z) &= \frac{1}{n_\textrm{tot}}\frac{dn}{dz} \\
    W_{\kappa}(z) &= \frac{3}{2}\Omega_mH_0^2\frac{(1+z)}{H(z)}\frac{\chi(z)}{c}\left(\frac{\chi(z_*)-\chi(z)}{\chi(z_*)}\right) \\
    W_y(z) &= \frac{1}{1+z},
\end{align}
where $\frac{dn}{dz}$ is the galaxy redshift distribution, $n_\textrm{tot}$ is the total number of galaxies in the sample, and $z_*$ is the redshift of the last scattering.  We neglected the magnification bias contribution to the galaxy kernel, which is more important for high-redshift galaxy samples. We show our derivation for the Compton-$y$ kernel in Appendix~\ref{sec:Cl_yy}, and more details on the CMB gravitational lensing kernel can be found in \citet{2001PhR...340..291B} and \citet{2006PhR...429....1L}.
\subsection{Halo model}\label{sec:halo_model}
In the halo model formalism \citep[for a review, see][]{2002PhR...372....1C}, all dark matter is assumed to reside in halos.
We defined a halo as the spherical region in which the average density is $\Delta = 200$ times the cosmic mean density. This region has a radius $r_{200\textrm{m}}$ , and the associated mass is $M_{200\textrm{m}}$. If not specified otherwise, we simply note $M_{200\textrm{m}}=M$. In the literature, halos are also defined as overdensities relative to the critical mass density, and several overdensity values are considered, mainly $\Delta=200$ or $\Delta=500$. Conversions between the different definitions require the adoption of a specific halo mass profile, as well as a concentration value.

We worked with $M_{500\textrm{c}}$ for the gas profile and converted between $M_{200\textrm{m}}$ and $M_{500\textrm{c}}$. In order to make this conversion, we followed \citet{2003ApJ...584..702H} and assumed a Navarro-Frenk-White (NFW) profile \citep{1997ApJ...490..493N} and a concentration following \citet{2004A&A...416..853D}, to remain consistent with the halo mass function derived by \citet{2008ApJ...688..709T}, who made use of this concentration-mass relation. The conversion was performed using
\begin{align}\label{eq:mass_conversion}
    M_{500\textrm{c}} = M_{200\textrm{m}}\frac{\Delta_{500\textrm{c}}(z)}{\Delta_{200\textrm{m}}}c_{200\textrm{m}}^{-3}x\left(\frac{\Delta_{500\textrm{c}}(z)}{\Delta_{200\textrm{m}}}f(c_{200\textrm{m}}^{-1})\right)^{-3}.
\end{align}
The quantities in this equation are defined in Appendix~\ref{sec:mass_conversion}. We probed the NFW profile, and therefore finding a non-NFW profile would require use of the observed profile when performing the mass conversion.

The matter distribution is then defined by the large-scale distribution of halos and the small-scale distribution inside individual halos. For the galaxy distribution, the approach is similar: we specified the spatial distribution of galaxies within halos and how galaxies populate these halos. The latter is given by the halo occupation distribution (HOD).

\subsubsection{HOD implementation}\label{sec:HOD_implementation}
Galaxies were split into centrals and satellites. Central galaxies reside at the center of their host halos, and satellites orbit around the centrals within the halo. We adopted an HOD parameterization following \citet{2005ApJ...633..791Z}. The mean numbers of central and satellite galaxies in a halo of mass $M$ are
\begin{align}
    \langle N_{\textrm c} \vert M\rangle &= \frac{1}{2}\left[1+\textrm{erf}\left(\frac{\log M-\log M_\textrm{min}}{\sigma_{\log M}}\right)\right]f_\textrm{inc}(M)\label{eq:Nc} \\
    \langle N_{\textrm s} \vert M\rangle &= \langle N_c \vert M\rangle\left(\frac{M-M_0}{M_1}\right)^{\alpha_g}\mathcal{H}(M>M_0)\label{eq:Ns},
\end{align}
where $\mathcal{H}$ is the Heaviside function, and $\textrm{erf}$ is the error function, defined as
\begin{align}
    \textrm{erf}(x) = \frac{2}{\sqrt{\pi}}\int_0^x e^{-t^2}dt.
\end{align}
The mean is taken over all halos of mass $M$. 

The $f_\textrm{inc}$ function was introduced by \citet{2015ApJ...806....2M} and enables us to take into account the fact that the CMASS sample is incomplete at low stellar masses. \citet{2016MNRAS.457.4021L} showed that CMASS is about $80\%$ complete at redshift $0.55$ at stellar mass $\log(M/M_\odot)=11.4$, while the completeness is very low at a low stellar mass. The $f_\textrm{inc}$ function is defined as
\begin{align}
    f_\textrm{inc}(M) = \textrm{max}(0,\textrm{min}(1+\alpha_\textrm{inc}(\log{M}-\log{M_\textrm{inc}}))),
    \label{eq:f_inc}
\end{align}
so that it is a function of the halo mass $M$ and takes values between $0$ and $1$. This function is defined such that the completeness is $1$ for masses $M>M_\textrm{inc}$. The completeness therefore increases when $M_\textrm{inc}$ decreases or when $\alpha_\textrm{inc}$ decreases.

However, \citet{2016MNRAS.457.4021L} noted that modeling the galaxy incompleteness with this function is only approximate because this approach assumes that the missing galaxies are a random selection at fixed halo mass and stellar mass, which is not the case because CMASS galaxies are color selected, and there is a correlation between galaxy colors and stellar masses. We leave evaluation of this effect for future work and refer to \citet{2016MNRAS.460.1457S} for an analysis focused on dealing with CMASS stellar mass incompleteness.

In order to reduce degeneracies, we restricted the number of free parameters by setting $M_0 = M_\textrm{min}$. Physically, this meant that we imposed that satellite galaxies started forming in halos with a mass high enough to have $0.5$ central galaxies on average. Nevertheless, this is not a strong condition because $M_0$ is difficult to constrain, but has little impact. We also fixed $\alpha_g = 1$, which is found by a number of studies \citep[e.g., ][]{2005ApJ...633..791Z,2011ApJ...736...59Z,2015ApJ...806....2M}. Since the HOD model we used is phenomenological, this means that we restricted our analysis to a subclass of HOD parameterization with less freedom. This, however, is a more important assumption that we discuss in Sect.~\ref{sec:HOD_discussion}. 

Additionally, when running our Monte Carlo Markov Chain (MCMC) pipeline (see Sect.~\ref{sec:results}) we added a Gaussian prior on $\log{M_\textrm{inc}}$ centered around $\log{M_\textrm{min}}$ with a width $\sigma=0.6,$ which is about twice the 1$\sigma$ found by \citet{2015ApJ...806....2M}. Adding this prior explicitely assumes that $\log{M_\textrm{inc}}$ should be on the same order of magnitude as $\log{M_\textrm{min}}$, which is motivated by the idea that low stellar mass galaxies (which are missing in CMASS) probably reside in low-mass halos whose limit we estimate roughly at $M_\textrm{min}$ (the value at which halos contain $0.5$ central galaxies on average). Finally, for clarity, we recall the free parameters in the HOD that were varied, which are $M_\textrm{min}$, $\sigma_{\log M}$, $M_1$, $\alpha_\textrm{inc}$ , and $M_\textrm{inc}$.

\subsubsection{Implementing the halo model}
The matter distribution is described as a sum of the one-halo and two-halo terms that describe the distribution of matter inside an individual halo and the distribution of halos in space, respectively. In harmonic space, we write 
\begin{align}
    P(k,z) = P^\textrm{1h}(k,z)+P^\textrm{2h}(k,z),
\end{align} where $P$ denotes the power spectrum.
We used a theoretical framework closely following the approach described in \citet{2013MNRAS.430..725V} and \citet{2013MNRAS.430..767C} to compute each term.

Specifically, \citet{2013MNRAS.430..725V} showed that these terms can be computed as
\begin{align}
    P_{xy}^\textrm{1h}(k,z) = \int dM H_x(k \vert M,z)H_y^*(k \vert M,z)n(M,z)
\end{align}
\begin{multline}\label{eq:P_2h}
    P_{xy}^\textrm{2h}(k,z) =
\int dM_1 H_x(k \vert M_1,z)n(M_1,z) \\
\times \int dM_2 H_y^*(k \vert M_2,z)n(M_2,z)Q(k\vert M_1,M_2,z),
\end{multline}
where $x$ and $y$ take values in \{m, c, s, $y$\}, standing for matter, central galaxies, satellite galaxies, and Compton-$y$ parameter, respectively. For galaxies, these expressions, with the $H_x$ functions described below, are valid assuming that the occupation numbers of centrals and satellites are independent and that the number of satellites follows a Poisson distribution \citep[for more details, see][]{2013MNRAS.430..725V}.

The halo mass function, $n$, is described in the next section, whereas in the simplest case, the function $Q$ reduces to $b(M_1)b(M_2)P^\textrm{lin}(M)$, where $b$ is the halo bias, and $P^\textrm{lin}$ refers to the linear-theory power spectrum. However, in order to account for the radial dependence of the halo bias and the halo exclusion effect, we employed the more detailed approach described in Appendix~\ref{sec:Q_k}.

\citet{2013MNRAS.430..725V} gave the functions $H_x$ for the  components
\begin{align}\label{eq:uh}
H_\textrm{m}(k \vert M,z) &= \frac{M}{\bar{\rho}_\textrm{m}}\tilde{u}_\textrm{h}, \\
\label{eq:uc}
H_\textrm{c}(k \vert M,z) &= \frac{\langle N_\textrm{c} \vert M\rangle}{\bar{n}_\textrm{g}(z)}, \\
\label{eq:us}
H_\textrm{s}(k \vert M,z) &= \frac{\langle N_\textrm{s} \vert M\rangle}{\bar{n}_\textrm{g}(z)}\tilde{u}_\textrm{s}(k \vert M,z),
\end{align}
with $\tilde{u}$ the Fourier transform of the normalized (meaning that its integral over real space is equal to $1$) radial profile of halo mass (subscript h) and satellite distribution (subscript s) (see Sect.~\ref{sec:gnfw}), and $\bar{n}_\textrm{g}(z)$ the comoving number density of galaxies at redshift $z,$ computed through
\begin{align}
    n_\textrm{g}(z) = \int \left[\langle N_\textrm{c} \vert M\rangle+\langle N_\textrm{s} \vert M\rangle\right] n(M,z)dM.
\end{align}

For the Compton-$y$ parameter, we added 
\begin{align}
    H_y(k\vert M,z) = \frac{c}{H}\tilde{y}(k\vert M,z),
\end{align}
where $\tilde{y}$ is the Fourier transform of the radial distribution $y(r\vert M)=\frac{\sigma_\textrm{T}}{m_\textrm{e}c^2}\Pe(r\vert M)$ such that
\begin{align}
    \tilde{y}(k\vert M) = \int_0^{\infty}4\pi r^2y(r\vert M)\frac{\sin(kr)}{kr}dr,
\end{align}
with $m_{\rm e}$ the electron mass and $\sigma_{\rm T}$ the Thomson scattering cross-section. In practice, the integral was performed from $r_\textrm{min}=10^{-6}r_{500\textrm{c}}$ to $r_\textrm{max}=50r_{500\textrm{c}}$. We used the pressure profile, $\Pe(r\vert M)$, from \citet{2010A&A...517A..92A} with the parameters taken from \citet{2013A&A...550A.131P},
\begin{multline}\label{eq:pressure_profile}
    \Pe(r\vert M) = 1.65(h/0.7)^2\,\textrm{eV}\,\textrm{cm}^{-3}\\
    \times E^{8/3}(z)\left(\frac{(1-\bh)M_{500c}}{3\times 10^{14}(0.7/h)M_\odot}\right)^{2/3+\alpha_p}p(r\vert (1-\bh)M_{500c}),
\end{multline}
where $p$ is the generalized Navarro-Frenk-White (GNFW) profile (see Sect.~\ref{sec:gnfw} for more details), $\alpha_p = 0.12$ and $E(z)=H(z)/H_0$. This profile was fit on clusters assuming that they are in hydrostatic equilibrium (HSE). Since this may not be the case, we converted the true halo mass into the effective hydrostatic mass with the $b_{\rm h}$ term, also known as the hydrostatic bias. This profile was calibrated on the $M_\textrm{500c}$ mass definition, so that we converted from the $M_\textrm{200m}$ used mostly in this work into $M_\textrm{500c}$ through Eq.~(\ref{eq:mass_conversion}).

The cross-correlation power spectra involving galaxies have contributions from central and satellite galaxies (which are not individually observable),
\begin{align}
    P_\textrm{gg}^\textrm{1h}(k,z) &= 2P_\textrm{cs}^\textrm{1h}(k,z)+P_\textrm{ss}^\textrm{1h}(k,z) \\
    P_\textrm{gg}^\textrm{2h}(k,z) &= P_\textrm{cc}^\textrm{2h}(k,z)+2P_\textrm{cs}^\textrm{2h}(k,z)+P_\textrm{ss}^\textrm{2h}(k,z) \\
    P_\textrm{gm}(k,z) &= P_\textrm{cm}(k,z)+P_\textrm{sm}(k,z) \\
    P_{\textrm{g}y}(k,z) &= P_{\textrm{c}y}(k,z)+P_{\textrm{s}y}(k,z),
\end{align}
where the one-halo and two-halo terms in the last two equations are identical, so that we do not specify them.

\subsubsection{Halo mass function and halo bias}\label{sec:hmf}
The halo mass function gives the comoving number of halos of mass $M$ per unit volume and logarithmic mass. In the halo model, we assumed that the halo mass function depends only on halo mass and redshift. We used the halo mass function that \citet{2008ApJ...688..709T} defined and calibrated on numerical simulations,
\begin{align}
    n(M,z) = \frac{\bar{\rho}_\textrm{m}}{M^2}\nu f(\nu) \frac{d\ln\nu}{d\ln M},
\end{align}
with 
\begin{align}
    f(\nu) = \alpha_t\left[1+(\beta_t\nu)^{-2\Phi_t}\right]\nu^{2\eta_t}e^{-\gamma_t\nu^2/2},
\end{align}
where \citet{2008ApJ...688..709T} found $\beta_t = \beta_{t,0}(1+z)^{0.20}$, \mbox{$\phi_t = \phi_{t,0}(1+z)^{-0.08}$}, $\eta_t = \eta_{t,0}(1+z)^{0.27}$, and $\gamma_t = \gamma_{t,0}(1+z)^{-0.01}$ with $\beta_{t,0} = 0.589$, $\gamma_{t,0}=0.864$, $\phi_{t,0}=-0.729$, $\eta_{t,0}=-0.243$. The value of $\alpha_t$ was chosen so that matter is not biased with respect to itself, which is imposed by
\begin{align}\label{eq:bias_normalization}
    \int b(\nu)f(\nu)d\nu=1,
\end{align}
with $b$ the halo bias function and $\nu$ the peak height, defined as
\begin{align}
    \nu(z) = \frac{\delta_\textrm{sc}(z)}{\sigma(M)}.
\end{align}
As pointed out by \citet{2013MNRAS.430..725V}, the critical threshold for spherical collapse derived by \citet{1997ApJ...490..493N} is well approximated in the case of a flat universe by
\begin{align}
\delta_\textrm{sc}(z) = 0.15(12\pi)^{2/3}\frac{\Omega_\textrm{m}(z)^{0.0055}}{D(z)}.
\end{align}

The remaining term required to compute the peak height, the linear matter variance on the
Lagrangian scale of the halo, is given by
\begin{align}\label{eq:matter_variance}
\sigma^2(M) = \frac{1}{2\pi^2}\int P_\textrm{mm}^\textrm{lin}(k,0)\tilde{W}^2(kR)k^2dk,
\end{align}
where $P_\textrm{mm}^\textrm{lin}$ is the linear matter power spectrum that we calculated using {\sl CLASS}\footnote{https://github.com/lesgourg/class\_public} \citep{2011JCAP...07..034B}, and $\tilde{W}$ is the Fourier transform of the real space top-hat filter, equal to 
\begin{align}
\tilde{W}(kR) = \frac{3\left(\sin kR -kR\cos kR\right)}{(kR)^3},
\end{align}
with $R = \left(3M/4\pi\bar{\rho}_\textrm{m}\right)^{1/3}$.

The halo distribution depends on halo mass, which is accounted for through the mass-dependent halo bias function, $b$. Numerical simulations show that massive halos tend to be more clustered than smaller halos, but with the constraint Eq.~(\ref{eq:bias_normalization}). We made use of the simulation-fitted halo bias provided by \cite{2010ApJ...724..878T},
\begin{align}
    b(\nu) = 1-\frac{A_t\nu^{a_t}}{\nu^{a_t}+\delta_\textrm{sc}(z=0)^{a_t}}+B_t\nu^{b_t}+C_t\nu^{c_t},
\end{align}
with
\begin{align*}
    A_t &= 1.0+0.24y_te^{-(4/y_t)^4},\\
    a_t &= 0.44y_t-0.88, \\
    B_t &= 0.183, \\
    b_t &= 1.5, \\
    C_t &= 0.019+0.107y_t+0.19e^{-(4/y_t)^4}, \\
    c_t &= 2.4, \\
    y_t &= \log{\Delta.}
\end{align*}

The halo mass function provided by \citet{2008ApJ...688..709T} was normalized at $z=0$ to account for all the mass in the Universe: $\int f(\nu) d\nu=1$. At higher redshifts,  Eq.~(\ref{eq:bias_normalization}) approximately imposes this property. However, with our framework, integrals were performed over mass rather than peak height, $\nu$. In practice, when integrating over mass, we may find $\int Mn(M)dM \neq \bar{\rho}_\textrm{m}$, violating this condition. This comes from the fact that even though $\sigma(M)$ tends to $+\infty$ as $M$ approaches $0$, the divergence is only logarithmic with $k$ so that Eq.~(\ref{eq:matter_variance}) varies very slowly with mass at the low-mass end.
As a result, we chose to slightly modify the value of $f$ at low $\nu$; more precisely, \citet{2008ApJ...688..709T} stated that the mass function is calibrated for $\sigma^{-1}>0.25$, which corresponds approximately to $M=10^{10} M_\odot$, whereas the behavior of the fitting function at lower masses is arbitrary. We therefore computed the mass integrals between $M_8=10^{8}M_\odot$ and $M_{16}=10^{16}M_\odot$ and modified the mass function below $M_{10}=10^{10}M_\odot$ so that $\int_0^{\nu(M_{10})}f(\nu)d\nu=\int_{\nu(M_8)}^{\nu(M_{10})}f(\nu)d\nu$. We also adapted the halo bias over this mass range so that it equaled its $f$-weighted mean. We checked that varying the mass boundaries has no significant impact on the resulting power spectrum.

\subsubsection{Generalized Navarro-Frenk-White profile}\label{sec:gnfw}
We use the GNFW profile to model the small-scale distribution of dark matter mass and satellite galaxies. This profile, proposed by \citet{2007ApJ...668....1N}, is a generalization of the NFW profile \citep{1997ApJ...490..493N},
\begin{align}\label{eq:gnfw}
    \rho(r \vert M) \propto \frac{1}{\left(r/\rs\right)^{\gamma}\left(1+(r/\rs)^\alpha\right)^{(\beta-\gamma)/\alpha}},
\end{align}
where $\rs$ is a characteristic radius related to $r_\textrm{200m}$ through the concentration parameter $c$ by $r_\textrm{200m} = c\rs$, and $r_\textrm{200m}$ is defined by $M_\textrm{200m}=(\frac{4\pi}{3})200\rho_\textrm{m}r_\textrm{200m}^3$. The GNFW profile reduces to the standard NFW profile when $\alpha = \gamma = 1$ and $\beta=3$. In order to compute the different functions $\tilde{u}$, we took the Fourier transform of Eq.~(\ref{eq:gnfw}), normalized to unity as required by Eqs.~(\ref{eq:uh})-(\ref{eq:us}),
\begin{align}
    \tilde{u}(k\vert M) = \int_0^{r_\textrm{200m}} 4\pi r^2 \frac{\sin(kr)}{kr}\frac{\rho(r\vert M)}{M}dr.
\end{align}
For the dark matter and galaxy distributions, we fixed $\alpha$ and $\gamma$ to unity and varied only $\beta$ in Eq.~(\ref{eq:gnfw}) to test whether the matter and satellite galaxies follow NFW profiles. This parameter is denoted $\beta_m$ ( $\beta_s$) when it relates to the matter (satellite) profile.

In the same way as for the mass conversion, we used the concentration parameter given by \citet{2004A&A...416..853D}. Even though our profiles are different from NFW (when $\beta \neq 3$), changing $\beta$ affects the profile so much that the value of the concentration has very little impact. The concentration we used was then
\begin{align}
    c(M,z) = \frac{c_0}{1+z}\left(\frac{M}{10^{14}h^{-1}M_\odot}\right)^{\alpha_c},
\end{align}
where $c_0=9.59$ and $\alpha_c=-0.102$. 

For the gas distribution, \citet{2013A&A...550A.131P} derived a GNFW profile for the gas pressure in Eq.~(\ref{eq:pressure_profile}),
\begin{align}
    p(r \vert M) = \frac{P_0}{\left(r/\rs\right)^{\gamma}\left(1+(r/\rs)^\alpha\right)^{(\beta-\gamma)/\alpha}},
\end{align}
with $\alpha=1.33$, $\beta=4.13$, $\gamma=0.31$, $P_0=6.41$, and using a concentration parameter $c=1.81$. 

We sought to probe the gas pressure in the vicinity of galaxies to determine whether it depends on stellar mass. We had the choice to either assume a hydrostatic mass bias (in Eq.~(\ref{eq:pressure_profile})) and vary the pressure normalization $P_0$, or to fix $P_0$ and measure the hydrostatic bias. The two approaches are not equivalent because varying the hydrostatic bias has an effect on the profile shape by changing the value of $\rs$ through $\rs(1-\bh)^{1/3}$. We chose the second option because this also enabled us to search for any stellar mass dependence of the hydrostatic bias (assuming that the pressure profile from \cite{2010A&A...517A..92A} is indeed universal).
As for the HOD, we recall the parameters introduced in this section that were varied for the profiles:  $\beta_m$ and $\beta_s$ in the matter and satellite profiles, and $b_h$, the hydrostatic bias.

\subsection{Galaxy-matter cross-correlation amplitude}
We added another parameter for the cross-correlation between CMB lensing convergence and galaxy maps, the cross-correlation amplitude $A$. This modifies Eq.~(\ref{eq:los_projection}) to
\begin{align}\label{eq:XC_amplitude}
    C_\ell^{\kappa \textrm{g}} = A\int \frac{dz}{c}\frac{H(z)}{\chi(z)^2}W_\kappa(z)W_\textrm{g}(z)P^{\kappa \textrm{g}}(k=\frac{\ell}{\chi(z)},z).
\end{align}
In theory, this parameter should be equal to unity, but many studies \citep{2016MNRAS.456.3213G,2017MNRAS.464.2120S,2021MNRAS.501.1481H,2021arXiv210602551S,2021MNRAS.501.6181K} have found different values. This parameter can be used to check for consistency, because a deviation from unity could indicate a possible failure either in the data processing or in the theoretical modeling. 

It might also indicate the presence of stochasticity \citep[e.g., ][]{1999ApJ...520...24D,2016MNRAS.456.3213G}, since stochasticity would change the relation between galaxy and matter overdensities, which is assumed to be statistically deterministic and parameterized through the galaxy bias. If a stochastic component were to break this deterministic relation, it would change the galaxy auto-power spectrum, but not the matter-galaxy cross correlation. The galaxy bias measured from the galaxy auto-power spectrum would differ from the one measured with the cross correlation with CMB lensing, and as a result, $A$ would capture this deviation.

Additionally, $A$ could also measure deviations from General Relativity, for instance, if the two Newtonian potentials entering the perturbed Friedmann-Lemaître-Robertson-Walker line element were to be different. This is the goal of the $\mu$, $\gamma$ parameterization \citep{2009PhRvD..79h3513Z} or of the $E_G$ statistics \citep{2007PhRvL..99n1302Z}.
We did not introduce this parameter for the cross correlation between the Compton-$y$ and galaxy maps because it would be degenerate with the hydrostatic mass bias, which already probes the amplitude of $C_\ell^{yg}$.

\section{Method and results}
\label{sec:results}
\subsection{Method}
With our theoretical model described in the previous section, we used the MCMC sampler \textit{emcee}\footnote{https://emcee.readthedocs.io/} \citep{2013PASP..125..306F} to fit our observations. As a summary, our considered observations are the galaxy auto power spectrum, $C_\ell^\textrm{gg}$ ($16$ observational points after multipole binning), the galaxy-CMB lensing convergence cross correlation, $C_\ell^{\kappa \textrm{g}}$ ($16$ observational points), and the galaxy-Compton-$y$ parameter cross-correlation, $C_\ell^{y\textrm{g}}$ ($12$ observational points). We also added the galaxy abundance as an observational constraint, which is approximately constant with redshift because we only selected galaxies with redshift $z\in [0.47,0.59]$ where the  observations are the most complete. In order to account for the slight variation in the abundance with redshift, we adopted a $20$\% uncertainty for the abundance constraint in the likelihood.

We varied the parameters $M_\textrm{min}$, $\sigma_{\log M}$, $M_1$, $\alpha_\textrm{inc}$, and $M_\textrm{inc}$ in the HOD model, adding a flat prior between $0$ and $1$ on $\alpha_\textrm{inc}$ and a Gaussian prior with $\sigma=0.6$ on $\log{M_\textrm{inc}}$ (see Sect.~\ref{sec:HOD_implementation} for the discussion of this prior). We also varied $\beta_\textrm{m}$ and $\beta_\textrm{s}$ in the matter and satellite profiles. However, as we show in Sect.~\ref{sec:lensing_discussion}, $\beta_\textrm{m}$ cannot be constrained with the current data and was therefore fixed to $3$ (the value for which the GNFW is simply an NFW) to avoid adding some small amount of noise in the fitting procedure. 

Finally, the hydrostatic bias, $\bh$, and the CMB lensing convergence-galaxy cross-correlation amplitude, $A$, were also varied. We recall that we fixed the cosmological parameters, for which we take the values from \citet{2020A&A...641A...6P} (TT, TE, EE+lowE+lensing+BAO): $H_0 = 67.66\,\textrm{km}\,\textrm{s}^{-1}\,\textrm{Mpc}^{-1}$, $\Omega_\textrm{b} h^2 = 0.2242$, $\Omega_\textrm{c} h^2=0.11933$, $\tau = 0.0561$, $n_\textrm{s}=0.9665,$ and $\sigma_\textrm{8} = 0.8102$. Even though some of the relations (mass-concentration relation, halo mass function, and bias, etc.) were derived with a slightly different cosmology, we assumed that they are still valid in the \Planck\  cosmology and that a slight difference would be absorbed in our phenomenological parameterization (in the HOD model, e.g.).

We fit each stellar mass threshold bin separately to reduce computation time, meaning that we did not use  cross correlations between the different mass bins. As a result, our observational data vectors consist of $45$ observational points, and we have eight free parameters. Assuming a  Gaussian likelihood,
\begin{align}
\ln{\mathcal{L}} = -\frac{1}{2}\left[\left(X(\theta)-X^\textrm{obs}\right)^TC^{-1}\left(X(\theta)-X^\textrm{obs}\right)+\frac{\left(n_\textrm{g}(\theta)-n_\textrm{g}^\textrm{obs}\right)^2}{\sigma_{n_\textrm{g}}^2}\right].
\end{align}
where $X^T$ denotes the transpose of a vector $X$, $\theta$ is the parameter vector, $X$ the concatenation of the three angular (cross-) power spectra, $C$ the covariance matrix, $n_\textrm{g}$ the galaxy abundance, and $\sigma_{n_\textrm{g}}$ the galaxy abundance uncertainty. This likelihood does not show the Gaussian prior added on $M_\textrm{inc}$. Using the MCMC sampler \textit{emcee}, we searched for the maximum of this likelihood.

\begin{figure}
   \resizebox{\hsize}{!}{\includegraphics{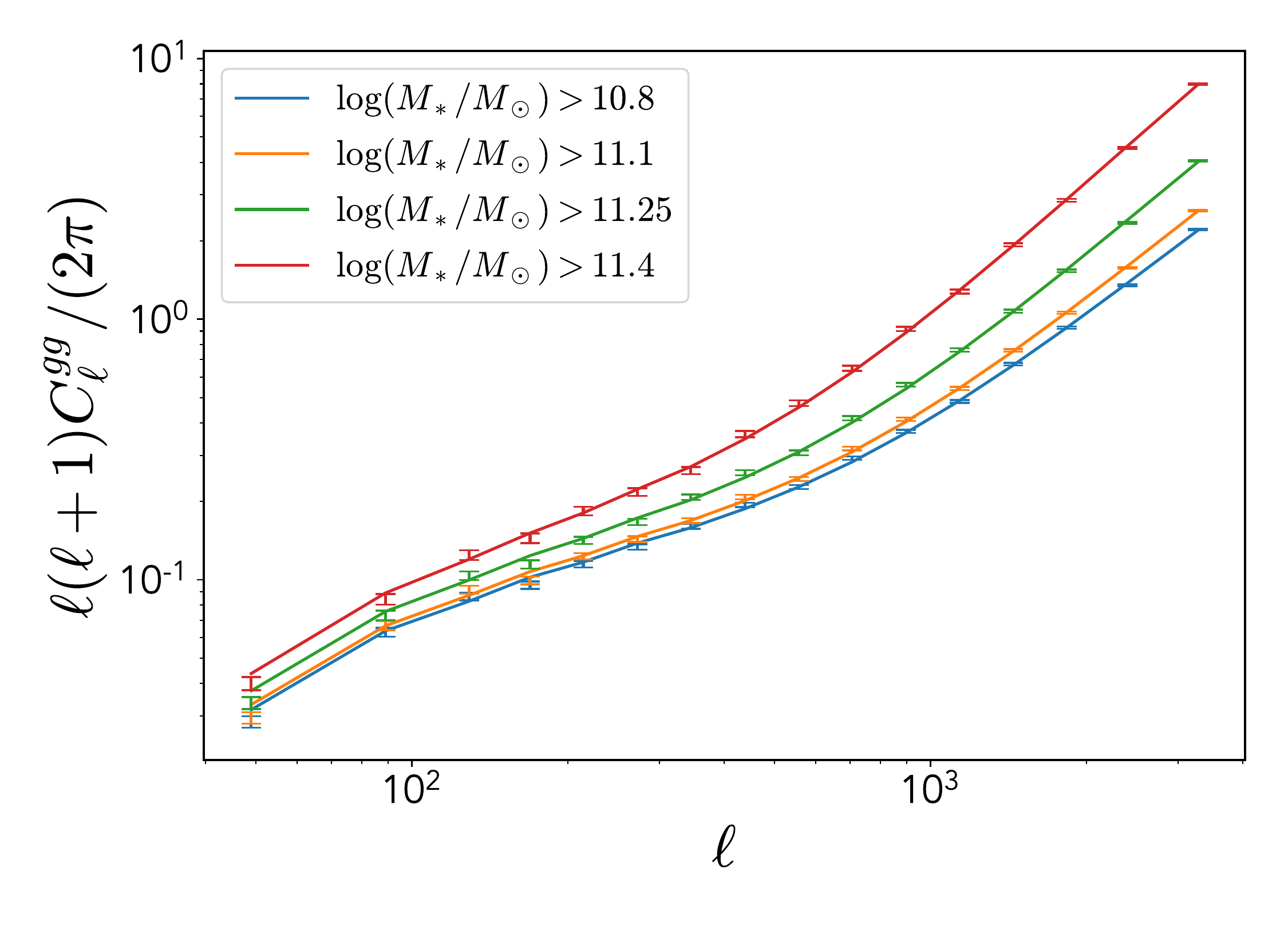}}
   \caption{Observed and best-fit galaxy angular auto-power spectra, $C_\ell^\textrm{gg}$, for the four different stellar mass threshold bins. The same colors are used for the observed error bars and the corresponding theoretical best fit in each stellar mass bin.}\label{fig:fit_cl_gg}
\end{figure}

\begin{figure*}
   \centering
   \includegraphics[width=17cm]{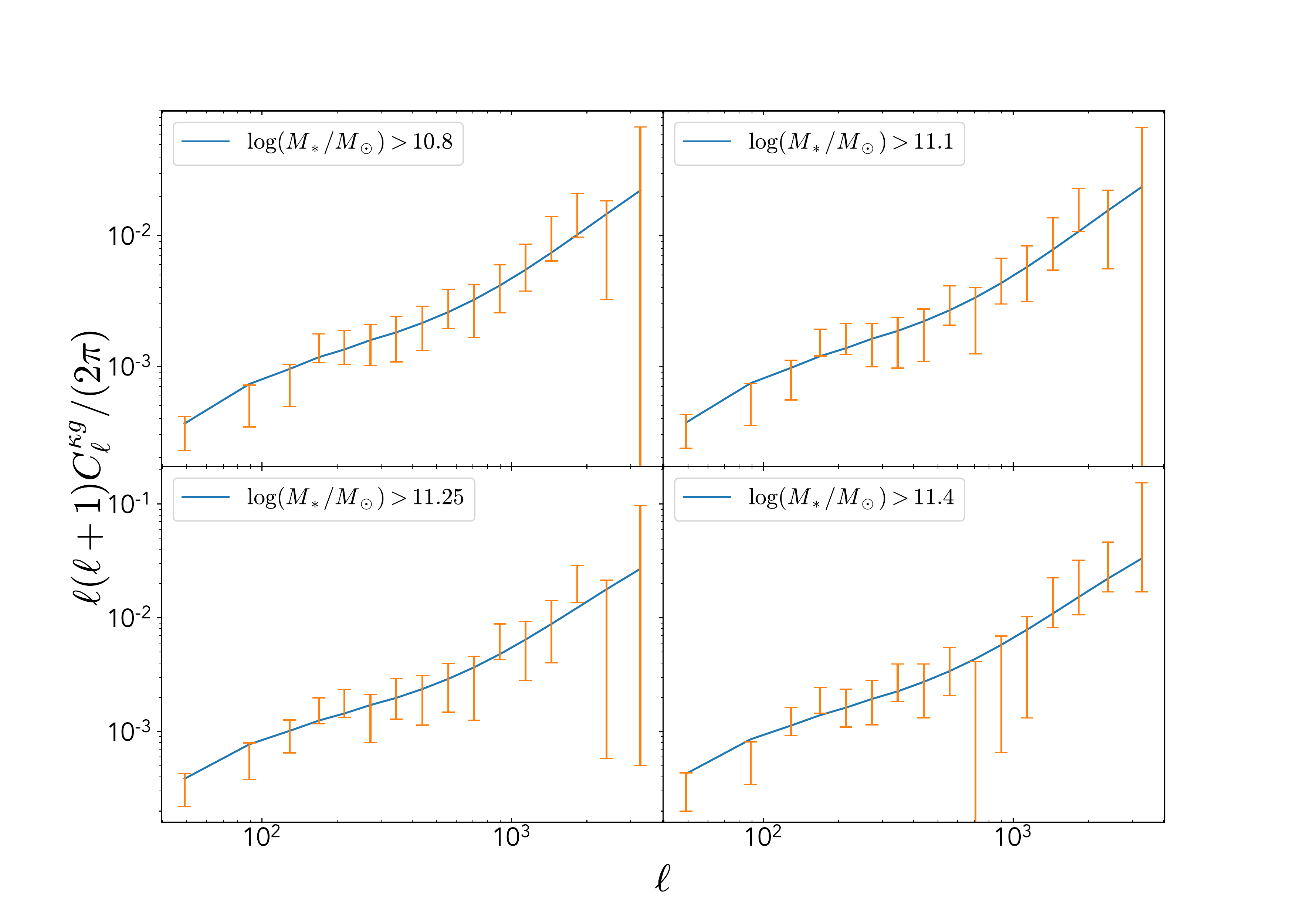}
   \caption{Observed and best-fit angular convergence-galaxy cross-power spectra, $C_\ell^{\kappa \textrm{g}}$, for the four different stellar mass threshold bins.}\label{fig:fit_cl_kg}
\end{figure*}

\begin{figure*}
   \centering
   \includegraphics[width=17cm]{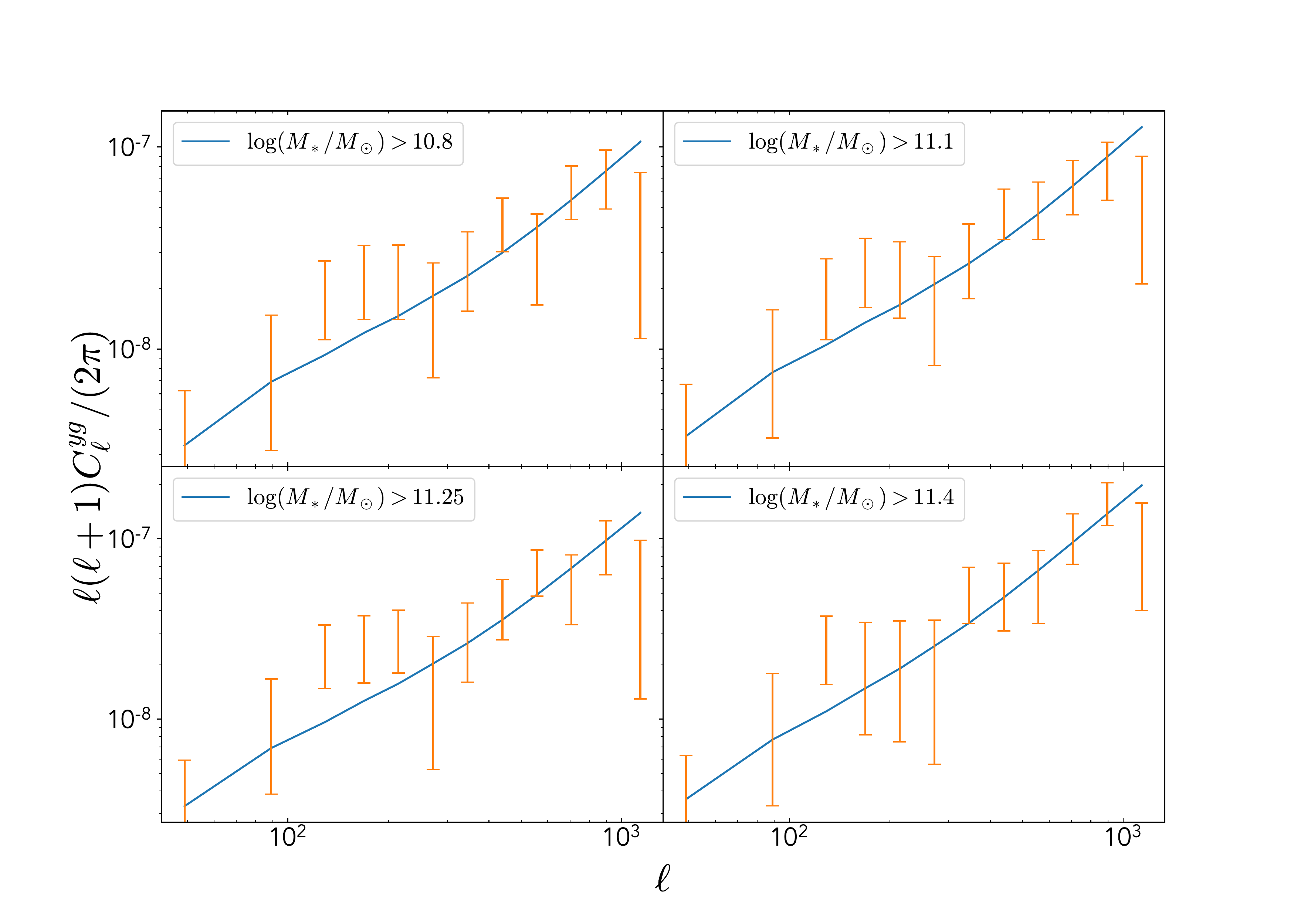}
   \caption{Observed and best-fit angular tSZ-galaxy cross-power spectra, $C_\ell^{y\textrm{g}}$, for the four different stellar mass threshold bins.}\label{fig:fit_cl_yg}
\end{figure*}

\subsection{Results}
\begin{figure*}
   \centering
   \includegraphics[width=17cm]{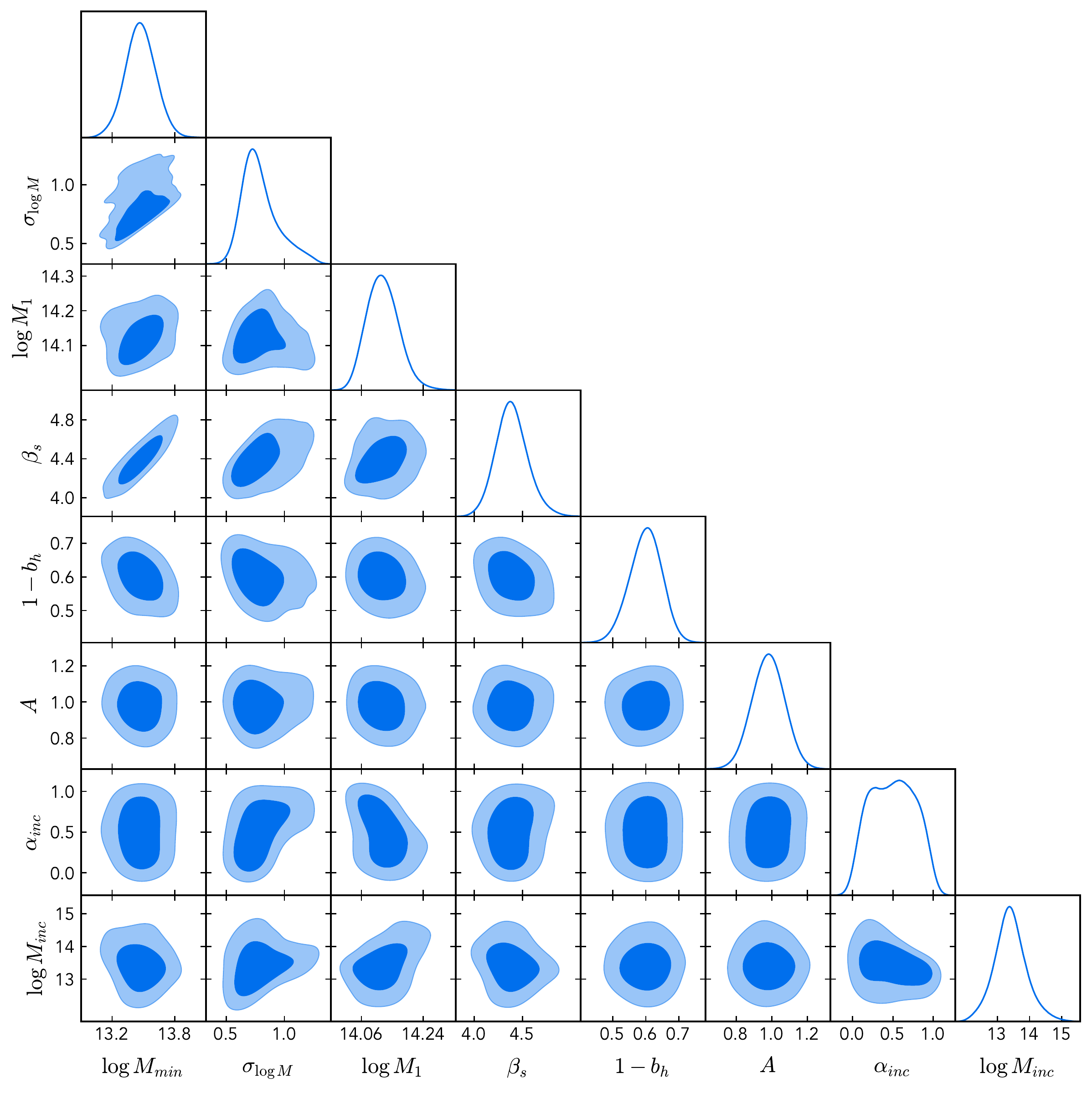}
   \caption{Posterior distributions  for the stellar mass threshold bin with  $\log(M_*/M_\odot)>10.8$. The $\beta_m$ parameter is not shown because it could not be constrained.}\label{fig:triangle_plot}
\end{figure*}

Figures~\ref{fig:fit_cl_gg}, \ref{fig:fit_cl_kg}, and \ref{fig:fit_cl_yg} show the theoretical (cross)-power spectra fitted to the observations for each stellar mass bin. The agreement is good overall, and we succeeded in reproducing the spectral amplitudes and shapes at both small and large scales. The goodness of the fits can be better evaluated with the reduced $\chi^2$ values given in Tab.~\ref{tab:estimated_parameters}. The table also gives the parameter medians together with the $68\%$ credible intervals, which are estimated via the $16th^\textrm{}$ and $84th^\textrm{}$ percentiles. We used the python package \textit{GetDist}\footnote{https://getdist.readthedocs.io/} \citep{2019arXiv191013970L} to show in Fig.~\ref{fig:triangle_plot} the degeneracies between the parameters for the less restrictive stellar mass threshold bin ($\log(M_*/M_\odot)>10.8$), except for the $\beta_\textrm{m}$ parameter, which we were not able to constrain (see Sect.~\ref{sec:lensing_discussion}).

\renewcommand{\arraystretch}{1.5}
\begin{table*}[ht]
    \center
    \begin{tabular}{|c|c c c c|}
    \hline
     & \multicolumn{4}{c|}{$\log(M_*/M_\odot)$} \\
    \hline
    Parameter & $>10.8$ & $>11.1$ & $>11.25$ & $>11.4$ \\
    \hline
    $\log{M_\textrm{min}}$ & $13.47^{+0.14}_{-0.14}$ & $13.58^{+0.14}_{-0.14}$ & $13.84^{+0.14}_{-0.13}$ & $14.20^{+0.12}_{-0.11}$ \\
    $\sigma_{\log{M}}$ & $0.76^{+0.19}_{-0.11}$ & $0.78^{+0.15}_{-0.10}$ & $0.86^{+0.0161}_{-0.079}$ & $0.959^{+0.148}_{-0.069}$\\
    $\log{M_1}$ & $14.119^{+0.049}_{-0.045}$ & $14.140^{+0.044}_{-0.044}$ & $14.171^{+0.047}_{-0.045}$ & $14.100^{+0.047}_{-0.055}$\\
    $\beta_\textrm{s}$ & $4.38^{+0.17}_{-0.15}$ & $4.71^{+0.16}_{-0.17}$ & $5.31^{+0.20}_{-0.17}$ & $6.35^{+0.24}_{-0.20}$\\
    $1-\bh$ & $0.602^{+0.046}_{-0.050}$ & $0.623^{+0.043}_{-0.050}$ & $0.558^{+0.046}_{-0.050}$ & $0.550^{+0.043}_{-0.049}$\\
    A & $0.981^{+0.091}_{-0.092}$ & $0.965^{+0.088}_{-0.075}$ & $0.956^{+0.092}_{-0.093}$ & $0.961^{+0.111}_{-0.110}$ \\
    $\alpha_\textrm{inc}$ & $0.51^{+0.31}_{-0.33}$ & $0.42^{+0.35}_{-0.28}$ & $0.39^{+0.33}_{-0.26}$ & $0.33^{+0.27}_{-0.21}$\\
    $\log{M_\textrm{inc}}$ & $13.39^{+0.49}_{-0.46}$ & $13.42^{+0.49}_{-0.48}$ & $13.69^{+0.58}_{-0.48}$ & $13.96^{+0.64}_{-0.54}$\\
    $\beta_\textrm{m, best fit}$ & $4.97$ & $5.91$ & $4.16$ & $10$ \\
    \hline
    $\chi^2/d.o.f$ & $1.190$ & $1.160$ & $1.512$ & $1.015$ \\
    \hline
    \end{tabular}
    \caption{Medians of the parameter posterior distributions and $68\%$ credible intervals. Because we found that $\beta_m$ could not be constrained, we fixed it to $3$ (hence recovering the NFW profile) and then fitted the other parameters. In this table, we report the value of $\beta_m$ corresponding to the maximum likelihood with all other parameters fixed to their best-fit values and that we denote $\beta_\textrm{m, best fit}$. The value of $\chi^2/d.o.f$ is also given for the best fit.}
    \label{tab:estimated_parameters}
\end{table*}

\subsubsection{Galaxy distribution}\label{sec:HOD_discussion}

Table~\ref{tab:estimated_parameters} shows that $M_\textrm{min}$ increases with the stellar mass, suggesting that galaxies with more stellar mass prefer to occupy more massive dark matter halos (see Eq.~\ref{eq:Nc}), as could be naively expected. At the same time, $\sigma_{\log{M}}$ also increases and is highly correlated with $M_\textrm{min}$ (see Fig.~\ref{fig:triangle_plot}). This is due to the fact that as minimum halo mass for a  central galaxy, $M_\textrm{min}$, increases, $\sigma_{\log{M}}$ can also increase to compensate for this, offsetting the impact on the mean number of central galaxies in lower-mass halos. This trend and correlation agree with the results found by \citet{2015ApJ...806....2M}. 

The parameters $\alpha_\textrm{inc}$ and $M_\textrm{inc}$, which were included to account for galaxy incompleteness at the low stellar mass end (see Eq.~\ref{eq:Nc} and Eq.~\ref{eq:f_inc}), are not well constrained and remain prior dominated, as shown in Fig.~\ref{fig:triangle_plot}. As a result, these parameters mainly act as a source of noise representing our lack of knowledge about the galaxies that are missing in the sample. In particular, we see degeneracies between $\sigma_{\log{M}}$, $\alpha_\textrm{inc}$ and $M_\textrm{inc}$; this comes from the fact that their effect on the power spectra is very similar. On the one hand, $M_\textrm{inc}$ and $\alpha_\textrm{inc}$ model the fact that some galaxies are missing in low-mass halos (meaning that these galaxies exist, but are not part of the sample), whereas on the other hand, $\sigma_{\log{M}}$ changes the number of galaxies in low-mass halos.

Figure~\ref{fig:fit_cl_gg} shows that  
clustering increases more rapidly at high multipoles for galaxies with high stellar mass (e.g., compare the slope of the red curve to the blue curve at high $\ell$). Since small-scale (high multipole) clustering is primarily driven through the one-halo term, this implies that massive galaxies tend to be more concentrated inside individual halos. In our model, this leads to an increase in $\beta_\textrm{s}$ (see Eq.~(\ref{eq:gnfw})), placing  satellite galaxies closer to their central galaxy. We indeed find that $\beta_\textrm{s}$ increases with stellar mass, and we conclude that higher-mass satellite galaxies have steeper radial profiles. This conclusion might change when $\alpha_g$ (see Eq.~(\ref{eq:Ns})) were allowed to vary because this also increases the clustering at small scales by raising the number of satellite galaxies. The two parameters are not completely degenerate, however, because $\alpha_g$ changes the total number of galaxies, in contrast to $\beta_\textrm{s}$. Because our HOD model is mainly constrained by the galaxy auto-power spectrum, we chose not to vary $\alpha_g$ to reduce the degeneracies between the parameters. 

\begin{figure}
   \resizebox{\hsize}{!}{\includegraphics{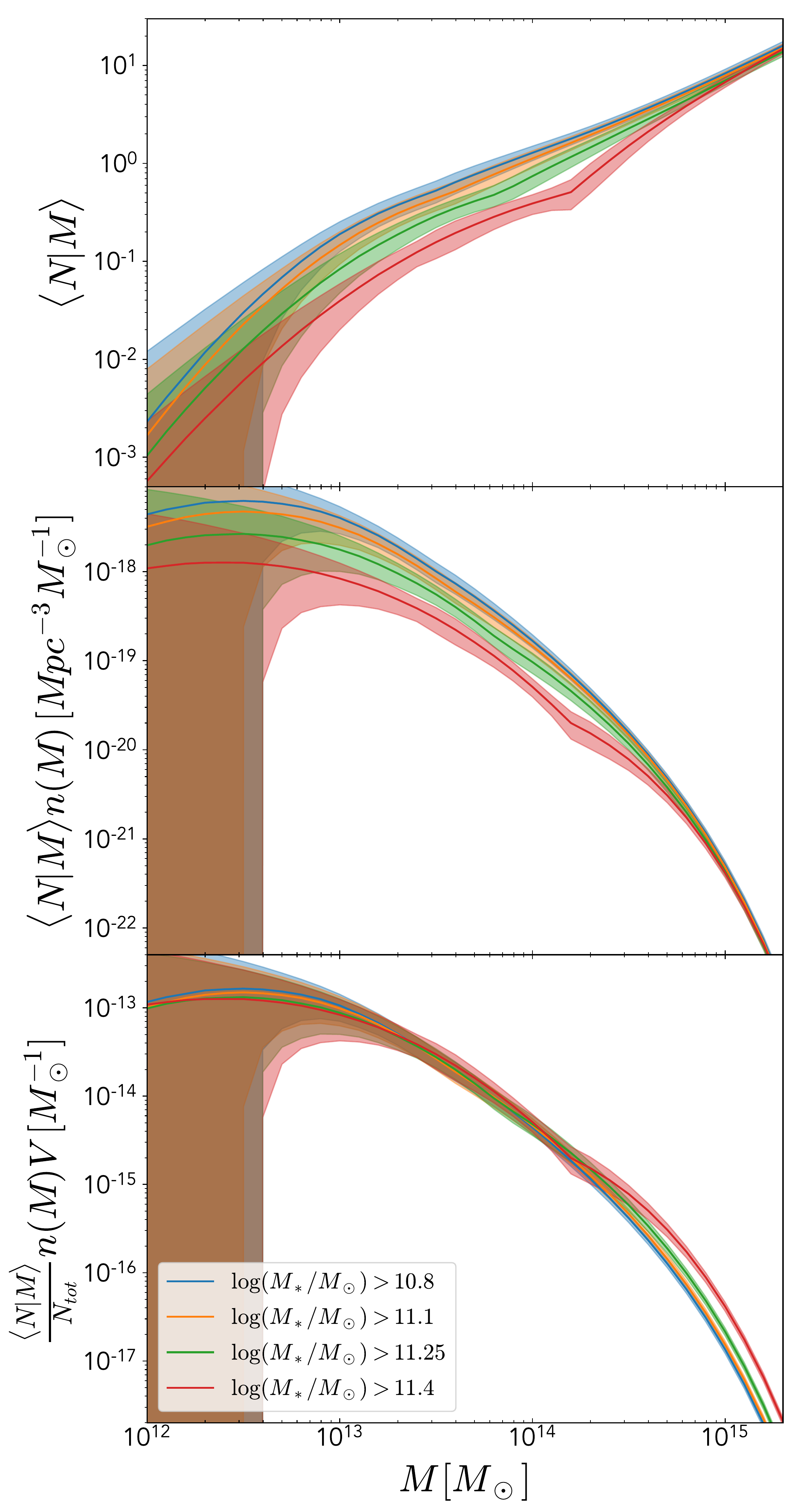}}
   \caption{Galaxy distribution as a function of the host halo mass. \textit{Top}: Total (central + satellite) number of galaxies contained in halos as a function of halo mass, $M$, evaluated at the sample median redshift. The four curves correspond to the different stellar mass threshold bins. The shaded bands represent the $1\sigma$ uncertainty. \textit{Middle}: Total number of galaxies per comoving volume and halo mass range (in units of $\textrm{Mpc}^{-3}M_\odot^{-1}$) as a function of halo mass.
   \textit{Bottom}: Total number of galaxies per halo mass range  normalized by the number of galaxies in each stellar mass bin. This fraction of galaxies lies in a halo of mass $M$. The quantity $V$ appearing in the label stands for the comoving volume.}\label{fig:N_tot_vs_M}
\end{figure}

The parameter $M_1$, which controls the number of galaxies at high halo mass (see Eq.~(\ref{eq:Ns})), is insensitive to stellar mass. 
This can be understood from the top and middle panels of Fig.~\ref{fig:N_tot_vs_M}, which show the number of galaxies inside individual halos as a function of halo mass. More galaxies of low stellar mass lie inside less massive halos (see the previous discussion on $M_\textrm{min}$), whereas the number of galaxies inside high-mass halos is approximately the same for each stellar mass bin. Similarly, the bottom panel of Fig.~\ref{fig:N_tot_vs_M} shows the proportion of galaxies lying in halos of different masses. There is a higher proportion of low stellar mass galaxies in low-mass halos and of high stellar mass galaxies in the most massive halos. The large uncertainties at low halo mass come from the poor constraints on $\alpha_\textrm{inc}$ and $M_\textrm{inc}$ and show that no galaxy in our sample may be residing in a halo with this mass.

\begin{figure}
   \resizebox{\hsize}{!}{\includegraphics{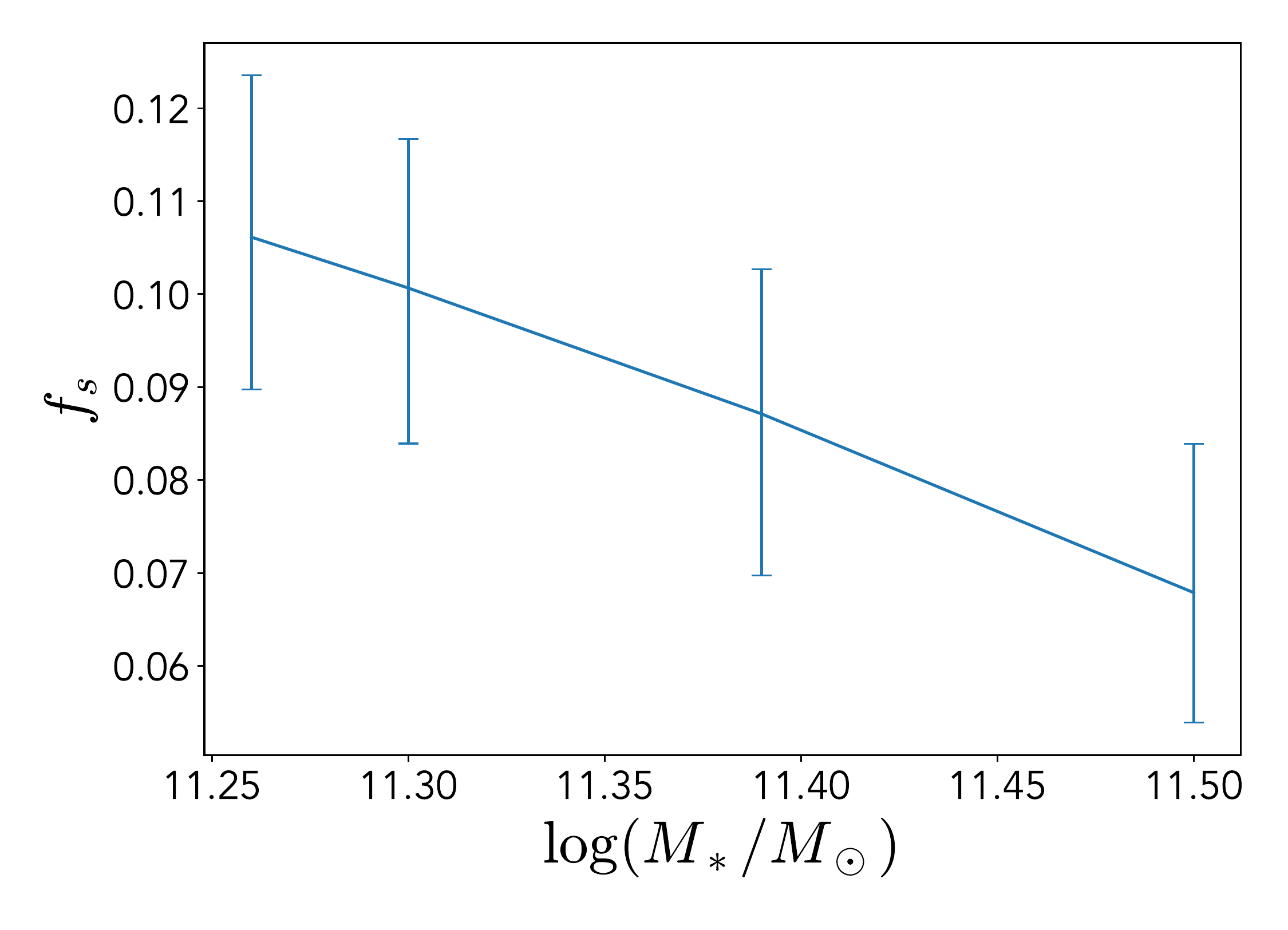}}
   \caption{Derived satellite fraction as a function of galaxy stellar mass. The error bars depict the $1\sigma$ credible intervals.}\label{fig:satellite_fraction}
\end{figure}

After constraining the HOD, we computed the satellite fraction in each subsample as
\begin{align}
    f_\textrm{s}(z) = \frac{\int dM \langle N_\textrm{s} \vert M \rangle n(M,z)}{\int dM \left[\langle N_\textrm{c} \vert M \rangle + \langle N_\textrm{s} \vert M \rangle\right] n(M,z)}.
\end{align}
The evolution of the satellite fraction with galaxy stellar mass is shown in Fig.~\ref{fig:satellite_fraction}. The satellite fraction decreases with stellar mass, so that the most massive galaxies tend to be centrals. This trend is expected, and is similar to what \citet{2015ApJ...806....2M} found, even though the values are slightly different. These differences may arise because the observations are not exactly the same: These authors also included galaxy weak lensing. Moreover, the satellite fraction is not an observable, but a derived parameter, so that modeling the differences can lead to different results. 

Making use of this HOD, we can also derive the galaxy linear bias describing the statistical relation between the galaxy distribution and the underlying dark matter field. The galaxy bias, $\bg$, is formally defined as $\delta_\textrm{g} = \bg \delta$, where $\delta_\textrm{g}$ and $\delta$ are the galaxy and matter overdensity fields, respectively. In the power spectrum formalism, this yields $P_\textrm{gg}(k)=\bg^2P_\textrm{mm}(k)$, so that using Eq.~(\ref{eq:P_2h}) in the large-scale linear limit leads to
\begin{align}
    \bg = \int dM \left[\langle N_\textrm{c} \vert M\rangle+\langle N_\textrm{s} \vert M \rangle\right]\frac{n(M,z)}{\bar{n}_\textrm{g}(z)}b(M,z).
\end{align}

\begin{figure}
   \resizebox{\hsize}{!}{\includegraphics{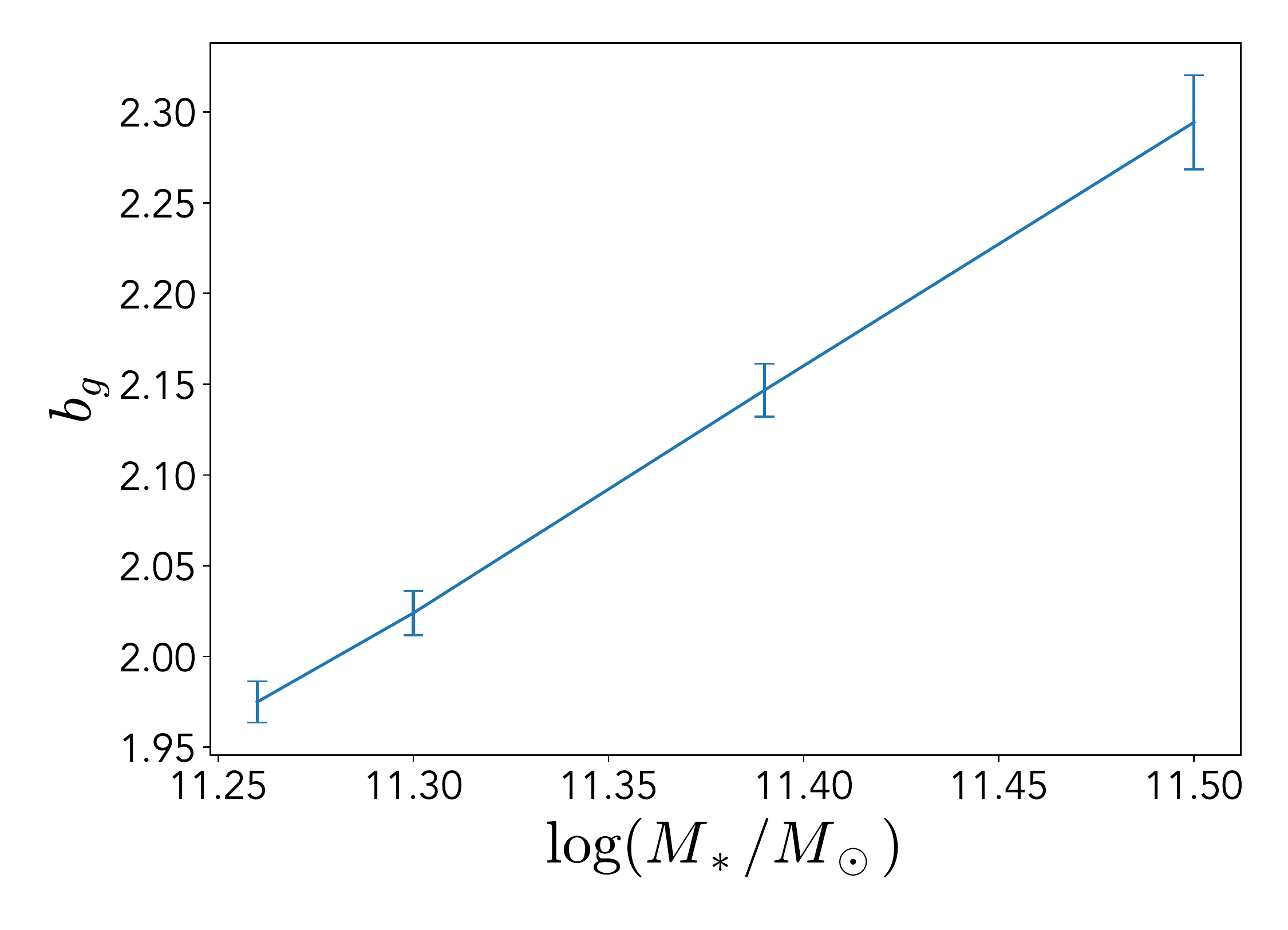}}
   \caption{Derived galaxy bias as a function of galaxy stellar mass. The error bars represent the $1\sigma$ credible intervals.}\label{fig:galaxy_bias}
\end{figure}

Figure~\ref{fig:galaxy_bias} shows the evolution of galaxy bias with stellar mass, together with the 1$\sigma$ interval. The increase in linear galaxy bias with stellar mass is clear, which is obvious from the galaxy power spectra in Fig.~\ref{fig:fit_cl_gg} (where the amplitude of the power spectra increases with galaxy stellar mass). This follows from our result that more massive galaxies populate higher-mass halos (e.g., Fig.~\ref{fig:N_tot_vs_M}), which are intrinsically more biased, that is, the halo bias increases with halo mass (this can be seen in the halo bias formula from \citet{2010ApJ...724..878T}, where the bias increases with $\nu$ and therefore with the halo mass). The important implication is that galaxy clustering depends on the properties of the galaxies under consideration; not all galaxies are interchangeable tracers of the dark matter field.

\subsubsection{Dark matter distribution}\label{sec:lensing_discussion}
The galaxy-CMB lensing convergence cross-correlation amplitude, $A$ (introduced in Eq.~\ref{eq:XC_amplitude}), is not sensitive to stellar mass, and its value is consistent with unity for each stellar mass bin. This supports our theoretical modeling as well as the data processing.  

\citet{2017MNRAS.464.2120S} found $A=0.78 \pm 0.13$ using the same data (\Planck\         CMB lensing and the CMASS survey, but without division into stellar mass bins) and working in configuration space. The difference between their result and ours may arise from our use of a restricted redshift range ($0.47<z<0.59$), from the use of different scale-binning schemes, and/or the fact that we have included much smaller scales in our study. This might indicate at an effect on large scales (either on the data processing or theoretical side) because Fig.~\ref{fig:fit_cl_kg} shows that our best-fit model falls in the upper part of the 1$\sigma$ interval on large scales (small multipoles). 

We therefore made another run for the first stellar mass bin, using only multipoles up to $\ell=492$ for the CMB lensing-galaxy cross-correlation. We obtained $A=0.86^{+0.13}_{-0.11}$ (compared to $0.981^{+0.091}_{-0.092}$ when using all the multipoles available in our analysis), so that we can confirm that $A$ indeed shifts toward lower values when only the largest scales are considered. This value is still consistent with unity. As mentioned previously, the remaining difference between our result and the result from \citet{2017MNRAS.464.2120S} might be due to the restricted redshift range and to differences in the scale binning schemes. The difference in our results is slight, however, when the uncertainties are taken into account.

The other parameter (apart from HOD parameters) used to fit the galaxy-CMB lensing cross-correlation power spectrum is $\beta_\textrm{m}$, which enters into the matter distribution profile parameterization (Eq.~\ref{eq:gnfw}). Unfortunately, the uncertainties are too large to quote a constraint on $\beta_\textrm{m}$, meaning that we cannot constrain the matter density profile even when $A$ is fixed to $1$. However, we examine in Sect.~\ref{sec:forecast} how well $\beta_\textrm{m}$ may be constrained by the future Simons Observatory CMB experiment.

\subsubsection{Gas distribution}

\begin{figure}
   \resizebox{\hsize}{!}{\includegraphics{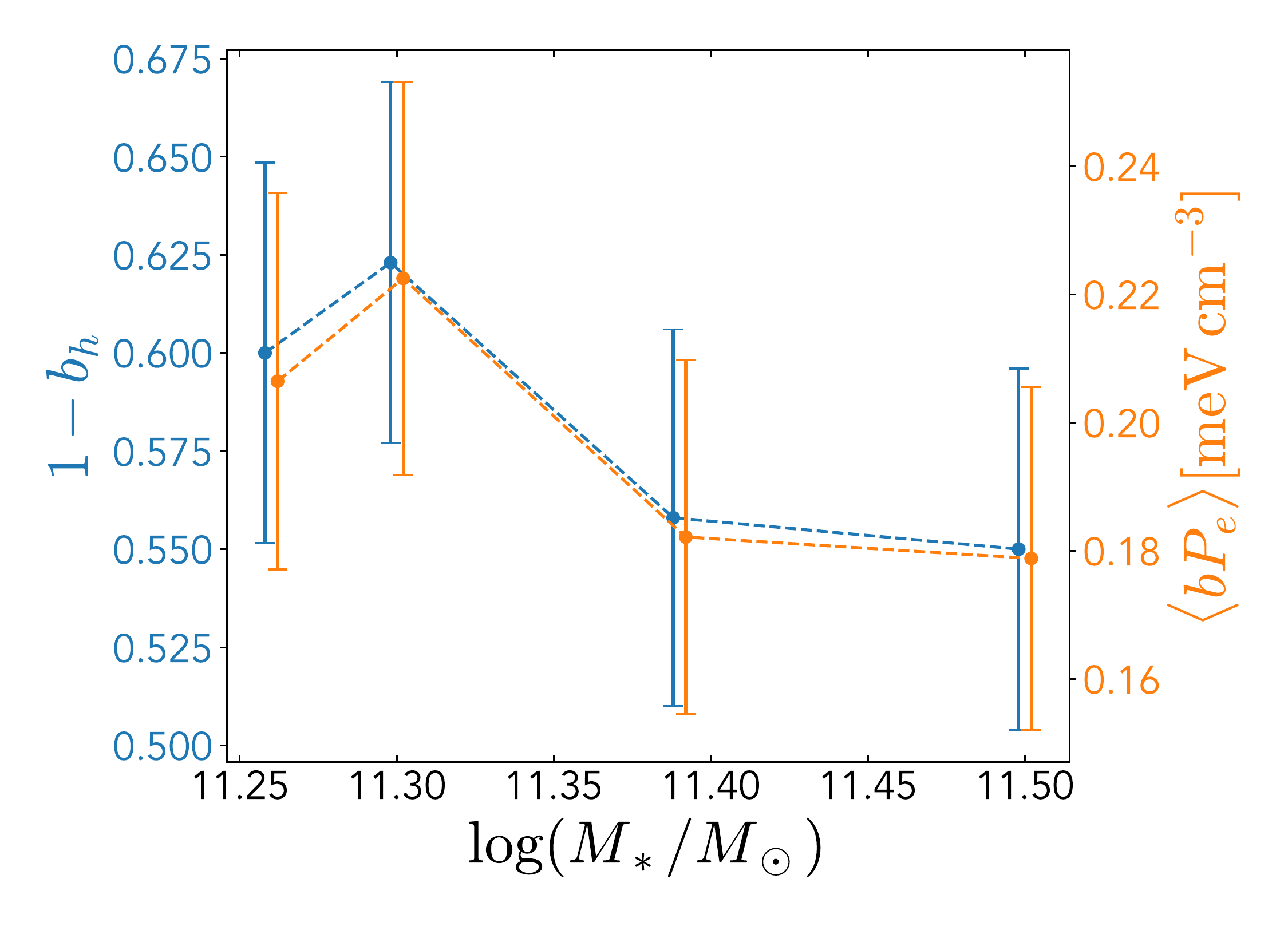}}
   \caption{Hydrostatic mass bias (blue) and halo-bias weighted thermal pressure (orange) as a function of galaxy median stellar mass. The halo-bias weighted thermal pressure is consistent with the results from \citet{2019PhRvD.100f3519P} when compared with the similar redshift bin. The two sets of error bars are shifted along the $x$-axis for visualization.} \label{fig:gas_bias}
\end{figure}

The Compton-$y$ cross-correlation probes the gas pressure in the vicinity of CMASS galaxies, and with our approach, this is quantified through the hydrostatic mass bias, $\bh$ (see Eq.~(\ref{eq:pressure_profile}). As shown in Tab.~\ref{tab:estimated_parameters} or in Fig.~\ref{fig:gas_bias}, this bias increases slightly with stellar mass, indicating that the gas pressure is lower around high stellar mass galaxies. We can also express the trend with the halo-bias weighted thermal pressure,  defined as
\begin{align}
    \langle b\Pe \rangle(z) = \int dM n(M,z)b(M,z) \int_0^{\infty} dr 4\pi r^2 \Pe(r\vert M,z),
\end{align}
where we recall that $b$ is the halo bias. This quantity is especially interesting because it is directly measured in the power spectrum on linear scales. This bias is also directly linked to the pressure profile integral, and because the halo bias increases around high stellar mass galaxies, finding a decreasing $\langle b\Pe\rangle$ means that the pressure profile integral must decrease with stellar mass. We note, however, that the data remain consistent with a constant value.

The dependence of $\langle b\Pe\rangle$ on stellar mass is also shown in Fig.~\ref{fig:gas_bias}, and the values found (approximately $0.18$ to $0.21\,\textrm{meV}\,\textrm{cm}^{-3}$) agree quite well with~\citet{2022PhRvD.105l3526P}, who performed the cross correlation between tSZ and galaxy shears.
The hydrostatic bias we infer takes values between $0.55$ and $0.62$, which also agrees very well with~\citet{2022PhRvD.105l3526P} ($1-\bh=0.56 \pm 0.03$) and is slightly lower than \citet{2018MNRAS.480.3928M} ($1-\bh=0.649 \pm 0.041$) and \citet{2022ApJ...935...18I} ($1-\bh=0.67 \pm 0.03$). 

As pointed out by \citet{2018A&A...614A..13S}, a mild difference is known on this parameter because tSZ number cluster counts combined with the observation of CMB primary anisotropies lead to low hydrostatic bias values of about $1-\bh = 0.6$, whereas estimates from weak-lensing and X-ray observations favor higher values of about $1-\bh = 0.8$-$0.85$. The hydrostatic mass bias we find is therefore consistent with CMB and tSZ cluster count observations \citep{Pclus2}. However, we emphasize that we kept the cosmology fixed, and allowing the cosmology to vary would broaden the error bars.

Assuming that the profile from~\citet{2010A&A...517A..92A} is universal and well calibrated, we find a hint that the hydrostatic bias increases (or $1-\bh$ decreases) with galaxy stellar mass. This is not statistically significant, but it is a tantalizing glimpse of the potential of this type of study to establish relations between galaxy properties and the gas surrounding them in the circumgalactic and intergalactic media. This may prove a powerful probe of galaxy feedback regulating the baryon cycle, galaxy evolution, and star formation. We explore this potential further in Sect.~\ref{sec:forecast}. Interestingly, some recent studies using cluster gas fractions \citep[][and references therein]{2019A&A...621A..40E,2022arXiv220412823W} also find indications that the hydrostatic mass bias increases with cluster mass.

\section{Future potential: Forecast}\label{sec:forecast}
\subsection{Forecast specifications}
We forecast how well future CMB experiments will be able to constrain the value of $\beta_\textrm{m}$, $A,$ and $\bh$ taking the Simons Observatory (SO) specifications as an example \citep{2019JCAP...02..056A}. To do this, we used our theoretical model to create fiducial galaxy auto-power spectrum, $C_\ell^\textrm{gg}$, cross-correlation galaxy--CMB convergence spectrum, $C_\ell^{\kappa \textrm{g}}$, and galaxy--tSZ cross-power spectrum, $C_\ell^{y\textrm{g}}$, all assuming an NFW profile for the matter distribution ($\beta_\textrm{m}=3$) and ideal galaxy-matter coefficient ($A=1$). 

We adopted the first stellar mass bin (of the CMASS sample), meaning that we kept the same $\frac{dn}{dz}$ (number of galaxies, $dn$, per redshift slice, $dz$) and that we used our rounded best-fit HOD parameters as well as the best-fit hydrostatic bias. In order to also form an idea of potential improvement by using a future galaxy survey experiment, we considered very approximate specifications of the Dark Energy Spectroscopic Instrument (DESI) \citep{2016arXiv161100036D}. Considering a sample at similar redshift, the specifications of the luminous red galaxy (LRG) sample can be approximated as a sky coverage of $14000$ square degrees, and a $\frac{dn}{dzd\textrm{deg}^2} \approx 1000$. Using the same redshift range as previously ($0.47<z<0.59$), we obtain approximately $1.7\cdot 10^6$ galaxies, even though the real LRG survey of DESI will be at slightly higher redshift. This is about $\text{four}$ times the number of galaxies we have in our first CMASS stellar mass bin. Because this is only a very rough approximation of the specifications of DESI, we call this case DESI-like, and refer to the cross-correlation between CMASS and SO as the CMASS x SO case, and between the DESI-like survey and SO as the DESI-like x SO case.

We adopted the forecast CMB lensing and tSZ noise curves that are publicly provided by the SO Collaboration \citep{2019JCAP...02..056A}. For the lensing, we considered the noise curve of the minimum variance estimator combining TT, TE, EE, EB, and TB measurements assuming no deprojection and the baseline sensitivity. For the tSZ noise curve, we used the baseline sensitivity curve without deprojection obtained via internal linear combination (ILC) component separation.

These noise curves give the covariance matrix through Eq.~(\ref{eq:covariance}). We assumed that the common sky fraction is the same as between CMASS and \Planck\ for the CMASS x SO case, even though in reality, SO will not cover the full CMASS (or DESI) area. Similarly, for the DESI-like x SO case, we adopted a sky coverage of $14000$ square degrees. Thus, our forecasts illustrate the potential of possible future combinations of CMB and galaxy surveys with ideal overlap, rather than specific predictions for the actual combinations of the planned SO and DESI surveys. 

We adapted the binning scheme to the noise curves, which are provided up to high multipoles ($\ell_\textrm{max} = 5000$ for the lensing and $\ell_\textrm{max}=7980$ for the tSZ). We therefore used multipoles between $\ell=30$ and $\ell=7980$ for the galaxy auto-power spectrum (making 21 logarithmically spaced bins), as well as for the cross correlation with the tSZ (using these high multipoles will require a thorough beam modeling and point source and CIB decontamination), and between $\ell=30$ and $\ell=5000$ for the cross correlation with the CMB lensing. The noise curves for multipoles lower than $\ell=80$ are not provided for the tSZ, and in the range $\ell \in [30,80],$ we used the noise from the \Planck\ experiment. 

We used the same pipeline as in our main study to fit our set of nine parameters. Here, we are primarily interested in $\beta_\textrm{m}$, $A$, and $1-\bh$ in the case CMASS x SO, to see how well the constraints will be improved with SO, and we do not expect much improvement on the HOD parameters because in this case, we used the same galaxy survey as in the main study of this paper. In the DESI-like x SO case, however, we are interested in potential constraints on the nine parameters and do expect some improvement on the HOD parameters due to the lower shot noise and larger sky coverage.

\begin{figure*}
   \centering
   \includegraphics[width=17cm]{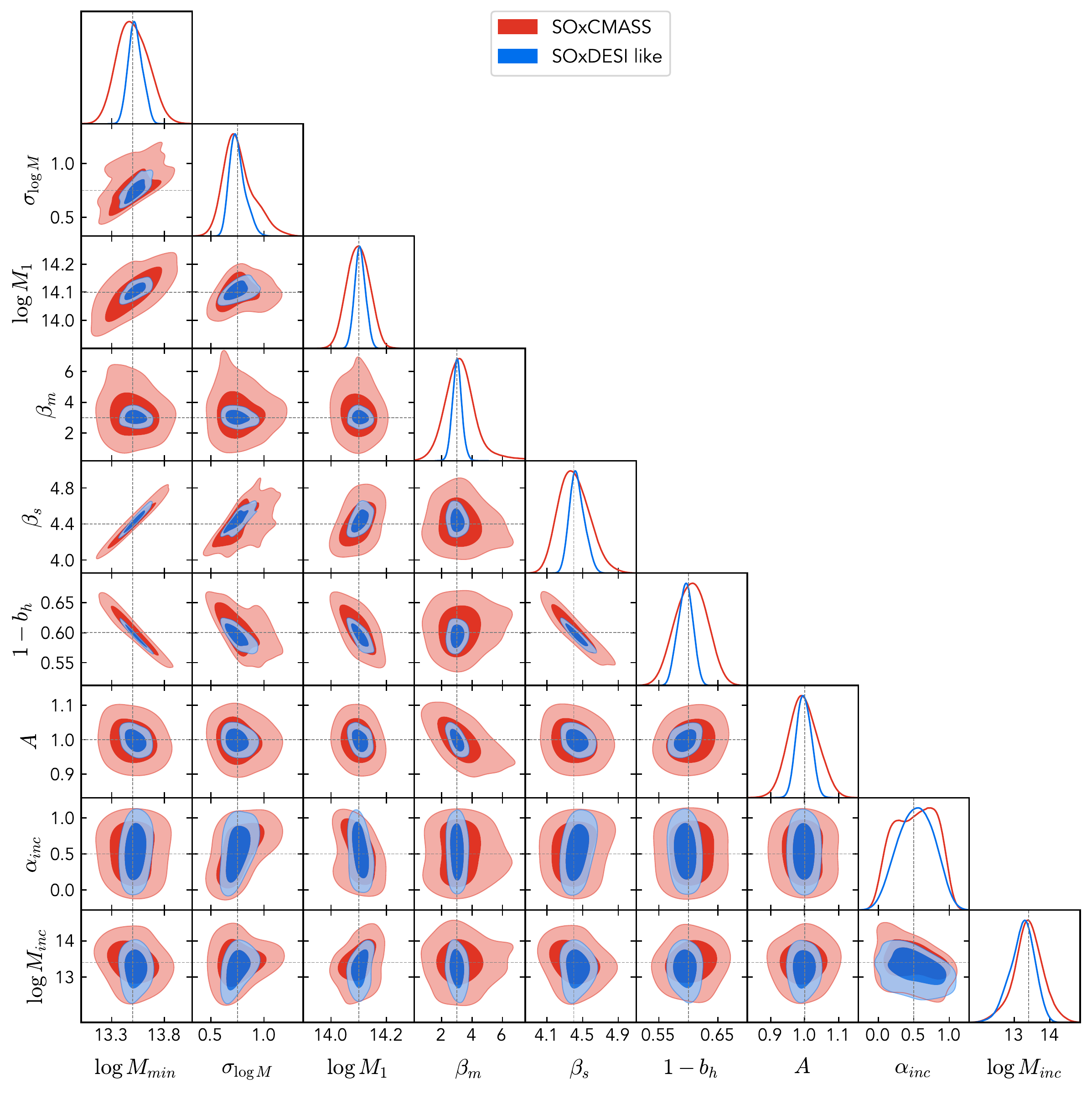}
   \caption{Forecast constraints with future experiments. \textit{Red}: Forecast constraints when the first stellar mass bin from a CMASS survey is cross correlated with full overlap with the future SO CMB experiment, using the noise curves provided by the SO collaboration. \textit{Blue}: Same, but using the galaxy shot noise and sky coverage of a DESI-like experiment with full overlap with SO (see  text for details). The dashed lines show the fiducial parameter values we used for the forecast.}\label{fig:triangle_plot_forecast}
\end{figure*}

Table~\ref{tab:forecast_parameters} and Fig.~\ref{fig:triangle_plot_forecast} show the constraints that can be achieved when we cross correlate the first stellar mass bin from CMASS and the DESI-like survey with SO over the aforementioned multipole ranges. Fig.~\ref{fig:triangle_plot_forecast} shows that the estimated parameters recover their fiducial values within 1$\sigma$, showing that the results are unbiased.

\renewcommand{\arraystretch}{1.5}
\begin{table*}[ht]
    \center
    \begin{tabular}{|c|c c c c|}
    \hline
    Parameter & Fiducial value & Forecast CMASS x SO 1$\sigma$ & Forecast DESI like x SO 1$\sigma$ & Current 1$\sigma$ with Planck data \\
    \hline
    $\log{M_\textrm{min}}$ & $13.5$ &  $0.14$ & $0.068$ & $0.14$\\
    $\sigma_{\log{M}}$ & $0.75$ & $0.14$ & $0.072$ & $0.15$ \\
    $\log{M_1}$ & $14.1$ & $0.041$ & $0.019$ & $0.047$\\
    $\beta_\textrm{m}$ & $3.00$ & $0.91$ & $0.30$ & $\infty$ \\
    $\beta_\textrm{s}$ & $4.4$ & $0.17$ & $0.083$ & $0.16$ \\
    $1-\bh$ & $0.6$ & $0.027$ & $0.013$ & $0.048$\\
    $A$ & $1$ & $0.042$ & $0.021$ & $0.091$\\
    $\alpha_\textrm{inc}$ & $0.5$ & $0.32$ & $0.27$ & $0.32$\\
    $\log{M_\textrm{inc}}$ & $13.4$ & $0.38$ & $0.35$ & $0.47$\\
    \hline
    \end{tabular}
    \caption{Fiducial values used for the input power spectra, with the 1$\sigma$ uncertainties obtained in the CMASS x SO and DESI-like x SO cases. The last column recalls the 1$\sigma$ uncertainties obtained in our study with \Planck\ for comparison. The 1$\sigma$ values reported here are the mean of the upper and lower values, even though both values are similar most of the time.}
    \label{tab:forecast_parameters}
\end{table*}

\subsection{Results in the CMASS x SO case}
In the case of the cross correlation between CMASS and SO, we find that the SO CMB lensing measurement will probe the matter distribution profile and constrain  $\beta_\textrm{m}$ at the level of $\sigma = 0.91$. Depending on how strongly $\beta_\textrm{m}$ is correlated with galaxy stellar mass, it may be possible to detect a trend. Furthermore, when $A$ was fixed to unity, we find $\sigma = 0.66$, implying that the constraint will be better if we can assume that galaxies are perfect tracers of dark matter (e.g., no stochasticity). 

The expected constraint on $1-\bh$ is nearly twice as strong as the constraint obtained with \Planck; the improvement arises from the lower tSZ noise and access to much higher multipoles.
Measuring this parameter to this precision would test the gas pressure dependence on stellar mass indicated in our study, and hence provide important information on feedback processes. Additionally, these higher multipoles would place constraints not only on the hydrostatic bias (i.e., the pressure profile normalization), but also on the  pressure profile. However, as mentioned previously, using these high multipoles will require mitigation of CIB and point source contamination. 

As expected, there is no significant improvement on the HOD parameter constraints. The reason is that the HOD parameters are mainly constrained by the galaxy auto-power spectrum because the cross-spectra already constrain other parameters.

\subsection{Results in the DESI-like x SO case}

In this case, the improvement on all parameters (except $\alpha_\textrm{inc}$ and $M_\textrm{inc}$) is significant.
The parameter $\beta_m$ entering the matter profile is constrained to $\sigma=0.30$, and even $\sigma=0.25$ when $A$ is fixed to unity. This case study shows the exciting potential of probing the small-scale matter distribution and its dependence on galaxy properties.

The hydrostatic bias constraint gains  another factor of $\text{two}$ compared to the CMASS x SO case. At this level, it would be possible to study feedback processes with high accuracy; for instance, relating gas thermal energy and mass to host active galactic nuclei (AGN) activity.

The constraints on the galaxy distribution parameters improve by a factor of two when compared to the CMASS x SO case. This comes from the much lower shot noise that  enables us to benefit from the smaller scales, as well as from the larger sky coverage.

This forecast shows that using cross correlations from SO and a DESI-like survey with full overlap would lead to improved understanding of the relations between dark matter, gas, and galaxies. It appears that correlating these data will enable researchers to study both the large- and small-scale distributions of all these components, a systematic cosmic census.

\section{Conclusion}
\label{sec:conclusion}
The advent of multiple wide-area surveys sensitive to galaxy properties, total mass, and gas mass enables a comprehensive study of all components of the cosmic web though cross-correlation studies. 
We examined the joint distribution of galaxies, dark matter, and gas as a function of galaxy stellar mass.  We divided CMASS galaxies into stellar mass bins and cross correlated them with the \Planck\ CMB lensing convergence and Compton-$y$ maps. Making use of the halo model, we fit an HOD and probed the large-scale structure of the Universe as well as the small-scale galaxy distribution through measurement of the galaxy and linear gas bias parameters and the galaxy halo distribution profile. We find in Sect.~\ref{sec:HOD_discussion} that the galaxy bias increases with stellar mass (see Fig.~\ref{fig:galaxy_bias}) because more massive galaxies preferably reside in higher-mass halos that are most strongly biased. We also find that most massive galaxies are central galaxies (see Fig.~\ref{fig:satellite_fraction}). Satellite galaxies tend to be less massive, and their radial distribution with their host halo seems to be steeper as their stellar mass increases. 

We measured the gas pressure around galaxies and find a hint that the pressure decreases slightly in the vicinity of the most massive galaxies, as shown in Fig.~\ref{fig:gas_bias}. The hydrostatic bias agrees  with previous studies, which also reported a slight increase with halo mass \citep{2019A&A...621A..40E,2022arXiv220412823W}. Finally, cross correlating the \Planck\ CMB lensing convergence map with galaxy maps, we measured a  cross-correlation amplitude (lensing parameter $A$) consistent with unity, implying that galaxies trace the dark matter field well; as a corollary, we found no evidence for stochasticity. However, the noise in the \Planck\ CMB lensing convergence map does not permit any constraint on the matter profile inside halos. In the near future, we forecast in Sect.~\ref{sec:forecast} that next generation CMB experiments will be able to do this by quantifying in particular expectations for SO as well as a DESI-like survey.

Cross-correlation studies such as the one presented here enable a comprehensive study of all components of the cosmic web, their interactions, and their coevolution. This is a powerful new technique with exciting potential to reveal details of large-scale cosmic structure and its emergence. 

\begin{acknowledgements}
      We thank Jean-Baptiste Melin for useful discussions about mass definitions conversions and gas pressure profiles. We acknowledge the use of the python libraries \textit{matplotlib} \citep{Hunter:2007}, \textit{numpy} \citep{harris2020array} and  \textit{scipy} \citep{2020SciPy-NMeth}. Some of the results in this paper have been derived using the {\sl healpy} and {\sl HEALPix} packages \citep{healpix}. 
\end{acknowledgements}

\bibliographystyle{aa} 
\bibliography{bibliography}

\begin{appendix}
\onecolumn
\section{Radius-dependent halo bias and halo exclusion}\label{sec:Q_k}

For our theoretical model to be accurate on intermediate and small scales, we must account for the radial dependence of halo bias and the effects of halo exclusion. These two effects are dealt with through $Q(k\vert M_1,M_2,z)$ in Eq.~(\ref{eq:P_2h}). We used the recipe given by \citet{2013MNRAS.430..725V} and \citet{2013MNRAS.430..767C}, which we summarize here for completeness.

We write the halo 2 pt correlation function, $\xi_{\rm hh}$, in terms of the matter 2 pt function, $\xi_{\rm mm}$, as
\begin{align}
    \xi_{hh}(r,z\vert M_1,M_2)=\left\{
    \begin{array}{ll}
        b(M_1,z)b(M_2,z)\zeta(r,z)\xi_{mm}(r,z) & \mbox{if } r>r_\textrm{min} \\
        -1 & \mbox{otherwise,}
    \end{array}
    \right.
\end{align}
where $\zeta(r,z)$ describes the radial dependence of the halo bias and $r_\textrm{min} = \textrm{max}(r_\textrm{vir}(M_1),r_\textrm{vir}(M_2))$ ensures the halo exclusion (i.e., that the center of a halo does not reside within another halo).
The radial bias function is given by 
\begin{align}
    \zeta(r,z)=\left\{
    \begin{array}{ll}
        \zeta_0(r,z) & \mbox{if } r \geq r_\psi \\
        \zeta_0(r_\psi,z) & \mbox{if } r \leq r_\psi,
    \end{array}
    \right.
\end{align}
where $r_\psi$ is defined through
\begin{align}
    \log{\bigg[\zeta_0(r_\psi,z)/\xi_{mm}(r_\psi,z)\bigg]}=\psi,
\end{align}
with $\psi=0.9$. The function $\zeta_0$ is fit to N-body simulations by \citet{2005ApJ...631...41T},
\begin{align}
    \zeta_0(r,z) = \frac{\left(1+1.17\xi_{mm}(r,z)\right)^{1.49}}{\left(1+0.69\xi_{mm}(r,z)\right)^{2.09}}.
\end{align}
\newpage
\section{Mass conversion}\label{sec:mass_conversion}
In this appendix, we derive Eq.~(\ref{eq:mass_conversion}) by applying the recipe described in \citet{2003ApJ...584..702H}.
The NFW profile has the form
\begin{align}
    \rho(r) = \frac{\rho_\textrm{s}}{\left(\frac{r}{\rs}\right)\left(1+\frac{r}{\rs}\right)^2}.
\end{align}
Integrating the mass profile until radii $r_{200\textrm{m}}$ and $r_{500\textrm{c}}$ gives
\begin{align}
    M_{200\textrm{m}} &= 4\pi\rho_\textrm{s} r_{200\textrm{m}}^3f(\rs/r_{200\textrm{m}})\label{eq:M200m}, \\
    M_{500\textrm{c}} &= 4\pi\rho_\textrm{s} r_{500\textrm{c}}^3f(\rs/r_{500\textrm{c}}) \label{eq:M500c},
\end{align}
where $f(x) = x^3\left(\ln(1+x^{-1})-(1+x)^{-1}\right)$. Using the mass definitions, we also have
\begin{align}
    M_{200\textrm{m}} &= \frac{4\pi r_{200\textrm{m}}^3}{3}\Delta_{200\textrm{m}}\rho_\textrm{m}(z) \label{eq:mass_def_M200m}, \\
    M_{500\textrm{c}} &= \frac{4\pi r_{500\textrm{c}}^3}{3}500\rho_\textrm{c}(z) \\
    &= \frac{4\pi r_{500\textrm{c}}^3}{3}\Delta_{500}(z)\rho_\textrm{m}(z) \label{eq:mass_def_M500c},
\end{align}
with $\Delta_{200\textrm{m}}=200$ and $\Delta_{500}(z) = 500\frac{\rho_\textrm{c}(z)}{\rho_\textrm{m}(z)}=\frac{500H(z)^2}{\Omega_\textrm{m} H_0^2 (1+z)^3}$.

Using Eqs.~(\ref{eq:M200m}) and (\ref{eq:M500c}), we have
\begin{align}
    \rho_\textrm{s} = \frac{M_{200\textrm{m}}}{4\pi r_{200\textrm{m}}^3f(\rs/r_{200\textrm{m}})} = \frac{M_{500\textrm{c}}}{4\pi r_{500\textrm{m}}^3f(\rs/r_{500\textrm{c}})}.
\end{align}
Plugging in the mass definitions, we can relate the two radii
\begin{align}
    &\frac{\Delta_{200\textrm{m}}\rho_\textrm{m}(z)}{3f(\rs/r_{200\textrm{m}})} = \frac{\Delta_{500}(z)\rho_\textrm{m}(z)}{3f(\rs/r_{500\textrm{c}})}, \\
    &f(\rs/r_{500\textrm{c}}) = \frac{\Delta_{500}(z)}{\Delta_{200\textrm{m}}}f(\rs/r_{200\textrm{m}}), \\
    &\frac{\rs}{r_{500\textrm{c}}} = x\left(\frac{\Delta_{500}(z)}{\Delta_{200\textrm{m}}}f(r_s/r_{200\textrm{m}})\right) = x\left(\frac{\Delta_{500}(z)}{\Delta_{200\textrm{m}}}f(c_{200\textrm{m}}^{-1})\right), \label{eq:radii_relation}
\end{align}
where $x$ is the reciprocal function of $f$. For completeness, we copy the fitting formula derived by \citet{2003ApJ...584..702H},
\begin{align}
    x(f) = \left(a_1f^{2p}+(3/4)^2\right)^{-1/2}+2f,
\end{align}
with $p=a_2+a_3\ln{f}+a_4(\ln{f})^2$ and $(a_1,a_2,a_3,a_4)=(0.5116,-0.4283,-3.13\times 10^{-3},-3.52\times 10^{-5})$.

Plugging Eq.~(\ref{eq:radii_relation}) into Eq.~(\ref{eq:mass_def_M500c}), we obtain
\begin{align}
M_{500\textrm{c}} = \frac{4\pi}{3}\Delta_{500}(z)\rho_\textrm{m}(z)r_s^3x\left(\frac{\Delta_{500}(z)}{\Delta_{200\textrm{m}}}f(c_{200\textrm{m}}^{-1})\right)^{-3}.
\end{align}
Finally, using $r_s = r_{200\textrm{m}}/c_{200\textrm{m}}$ and Eq.~(\ref{eq:mass_def_M200m}) to express $r_{200\textrm{m}}=~\left(\frac{3M_{200\textrm{m}}}{4\pi\Delta_{200\textrm{m}}\rho_\textrm{m}(z)}\right)^{1/3}$, we find
\begin{align}
    M_{500\textrm{c}} = M_{200\textrm{m}}\frac{\Delta_{500}(z)}{\Delta_{200\textrm{m}}}c_{200\textrm{m}}^{-3}x\left(\frac{\Delta_{500}(z)}{\Delta_{200\textrm{m}}}f(c_{200\textrm{m}}^{-1})\right)^{-3}.
\end{align}
\newpage
\section{Compton-$y$ parameter angular power spectrum}\label{sec:Cl_yy}
In this appendix, we derive the expression for the Compton-$y$ angular power spectrum  and its kernel. We first establish a result on the halo distribution that will prove useful in the derivation. 

We model the halo number density field, $n(M,\vec{r})$, as a Poisson realization, with mean $\bar{n}(M,\vec{r})$, of the underlying continuous matter overdensity field, $\delta(\vec{r})$. Taking  expectations over just the Poisson ensemble, $\langle \cdot \rangle_{\rm Poisson}$, for fixed $\delta(r)$, we have
\begin{align}
        \langle n(M,\vec{r}) \rangle_{\rm Poisson} &= \bar{n}(M,\vec{r}) = \bar{n}(M,z)(1+b(M)\delta(\vec{r})), \\
    \langle
    n(M_1,\vec{r}_1)n(M_2,\vec{r}_2) \rangle_{\rm Poisson} &= \bar{n}(M_1,\vec{r}_1)\bar{n}(M_2,\vec{r}_2)+\bar{n}(M_1,\vec{r}_1)\delta_D(\vec{r}_1-\vec{r}_2)\delta_D(M_1-M_2),
    \label{eq:poisson_variance}
\end{align}
where $b(M)$ is the halo bias and $\bar{n}(M,z)$ is the average number density of halos of mass $M$ at redshift $z$. In this appendix, $\bar{n}(M,z)$ is the halo mass function, which was denoted simply $n$ in the main body of the paper. In the above, $\delta_{\rm D}$ is the Dirac delta function. Taking the full expectation, $\langle\langle \cdot \rangle\rangle = \langle\langle\cdot\rangle_{\rm Poisson}\rangle$, over both the Poisson ensemble, using the above equations, and the $\delta(\vec{r})$-field ensemble, $\langle \cdot \rangle$, we find
\begin{align}
    \langle\langle n(M_1,\vec{r}_1)n(M_2,\vec{r}_2) \rangle\rangle
    &= \langle \bar{n}(M_1,\vec{r}_1)\bar{n}(M_2,\vec{r}_2)+\bar{n}(M_1,\vec{r}_1)\delta_D(\vec{r}_1-\vec{r}_2)\delta_D(M_1-M_2) \rangle \\ &= \langle\bar{n}(M_1,z_1)\left[1+b(M_1)\delta(\vec{r}_1)\right]\bar{n}(M_2,z_2)\left[1+b(M_2)\delta(\vec{r}_2)\right]+\bar{n}(M_1,z_1)\left[1+b(M_1)\delta(\vec{r}_1)\right]\delta_D(\vec{r}_1-\vec{r}_2)\delta_D(M_1-M_2) \rangle \\
    &\label{eq:xi_halos}= \bar{n}(M_1,z_1)\bar{n}(M_2,z_2)+b(M_1)b(M_2)\bar{n}(M_1,z_1)\bar{n}(M_2,z_2)\langle \delta(\vec{r}_1)\delta(\vec{r}_2) \rangle+\bar{n}(M_1,z_1)\delta_D(\vec{r}_1-\vec{r}_2)\delta_D(M_1-M_2).
\end{align}

The following derivation of the Compton-$y$ angular power spectrum is inspired by \citet{2011MNRAS.418.2207T}, who adopted the approach introduced by \citet{2006ApJ...643..598H}. The Compton-$y$ parameter is equal to the line-of-sight integral of the electron pressure,
\begin{align}
    y &= \frac{\kB\sigmaT}{\me c^2}\int dl\, n_\textrm{e} T_\textrm{e} \\
    &= \int adr\sum_{i=1}^Ny_\textrm{3D,c}(\vec{w}_i)u(\vec{w}_i-\vec{r}),
\end{align}
where $r$ is the comoving distance along the line of sight, $N$ is the number of clusters between $z=0$ and $z=z_\textrm{CMB}$, $\vec{w}_i$ is the comoving position of cluster $i$, $y_\textrm{3D,c}(\vec{w})$ is the value of $\frac{\kB T_\textrm{e}\sigmaT n_\textrm{e}}{\me c^2}$ at the center of cluster $i$, $u$ is the electron pressure profile, and $a$ is the scale factor. Changing the discrete sum to an integral over comoving position,
\begin{align}
    y &= \int adr \int dM\, y_\textrm{3D,c}(z,M)\int d^3w\, n(\vec{w},M)u(\vec{w}-\vec{r},M), 
\end{align}
where the value of $y_\textrm{3D,c}$ is assumed to only depend on the halo mass $M$ and on redshift $z(r)$. 

This expression can be written as a convolution, which reduces to a product in Fourier space,
\begin{align}
    y &= \int adr\int dM\, y_\textrm{3D,c}(z,M)\int \frac{d^3k}{(2\pi)^3}\, \tilde{n}(\vec{k},M)\tilde{u}(\vec{k},M)e^{i\vec{k}\cdot\vec{r}},
\end{align}
where $\tilde{x}=\mathcal{F}(x)$ denotes the Fourier transform of $x$. Using $e^{i\vec{k}\cdot \vec{x}}=4\pi\sum_{\ell,m}i^\ell j_\ell(kx)Y_{\ell m}^*(\hat{\vec{k}})Y_{\ell m}(\hat{\vec{x}})$, where the hat denotes a unit vector, we obtain the spherical harmonic decomposition
\begin{align}
    y_{\ell m} = \frac{i^\ell}{2\pi^2}\int adr\int dM \int d^3k\,\tilde{n}(\vec{k},M)\tilde{y}_\textrm{3D}(\vec{k},M,z)j_\ell(kr)Y_{\ell m}^*(\hat{\vec{k}}),
\end{align}
where $\tilde{y}_\textrm{3D}=y_\textrm{3D,c}\tilde{u}$. 
The angular power spectrum is then 
\begin{multline}\label{eq:cl_yy_xi_halos}
    \langle\langle y_{\ell m}y_{\ell m}^*\rangle\rangle = \frac{1}{4\pi^4}\int adra'dr'dMdM'd^3kd^3k'\, \tilde{y}_\textrm{3D}(\vec{k},M,z)\tilde{y}_\textrm{3D}^*(\vec{k}',M',z') \\
    \times j_\ell(kr)j_\ell(k'r')Y_{\ell m}^*(\hat{\vec{k}})Y_{\ell m}(\hat{\vec{k}}')\langle\langle\tilde{n}(\vec{k},M,z)\tilde{n}^*(\vec{k}',M',z')\rangle\rangle.
\end{multline}
We recall that the redshift is a function of comoving distance along the line of sight, $z(r)$.
The expectation at the end gives
\begin{align}
    &\langle\langle\tilde{n}(\vec{k},M,z)\tilde{n}^*(\vec{k}',M',z')\rangle\rangle = \langle\langle\mathcal{F}(n)(\vec{k},M,z)\mathcal{F}(n)(\vec{k}',M',z')^*\rangle\rangle \\
    &= \langle\langle\int d^3r\, n(M,r,z)e^{-i\vec{k}\cdot\vec{r}}\int d^3r'\, n(M',r',z')e^{i\vec{k}'\cdot\vec{r}'}\rangle\rangle \\
    &= \int d^3r\, e^{-i\vec{k}\cdot\vec{r}}\int d^3r'\, e^{i\vec{k}'\cdot\vec{r}'}\langle\langle n(M,r,z)n(M',r',z')\rangle\rangle.
\end{align}
With Eq.~(\ref{eq:xi_halos}),
\begin{multline}
    \langle\langle\tilde{n}(\vec{k},M,z)\tilde{n}^*(\vec{k}',M',z')\rangle\rangle = \int d^3r\, e^{-i\vec{k}\cdot\vec{r}}\int d^3r'\, e^{i\vec{k}'\cdot\vec{r}'}\bigg[\bar{n}(M,z)\bar{n}(M',z') \\
    +\bar{n}(M,z)\bar{n}(M',z')b(M,z)b(M',z')\langle \delta(\bold{r})\delta(\bold{r}')\rangle+\bar{n}(M,z)\delta_D(\bold{r}-\bold{r}')\delta_D(M-M')\bigg]
\end{multline}
\begin{multline}\label{eq:sum_3_terms}
    = (2\pi)^6\bar{n}(M,z)\bar{n}(M',z')\delta_D(\vec{k})\delta_D(\vec{k}')+(2\pi)^3\bar{n}(M,z)\bar{n}(M',z')b(M,z)b(M',z')D(z)D(z')P(\vec{k},z=0)\delta_D(\vec{k}-\vec{k}') \\
    +(2\pi)^3\bar{n}(M,z)\delta_D(\vec{k}-\vec{k}')\delta_D(M-M'),
\end{multline}
where $D(z)$ is the linear growth function.

Injecting each of these three terms into Eq.~(\ref{eq:cl_yy_xi_halos}), we show that the first makes no contribution, while the second term yields the two-halo power spectrum, and the third gives the one-halo contribution. The first term leads to
\begin{align}
    \langle\langle y_{\ell m}y_{\ell m}^*\rangle\rangle_1 = 16\pi^2\int adra'dr'dMdM'\,\tilde{y}_\textrm{3D}(0,M,z)\tilde{y}_\textrm{3D}^*(0,M',z')j_\ell(0)j_\ell(0)Y_{\ell m}^*(\hat{0})Y_{\ell m}(\hat{0})\bar{n}(M,z)\bar{n}(M',z').
\end{align}
Since $j_\ell(0)=0$ when $\ell>0$, this term is null.
The second term (which is the two-halo contribution) is
\begin{multline}
    \langle\langle y_{\ell m}y_{\ell m}^*\rangle\rangle_2 = C_\ell^{yy,2h} = \frac{2}{\pi}\int adra'dr'dMdM'd^3k\,\tilde{y}_\textrm{3D}(\vec{k},M,z)\tilde{y}_\textrm{3D}^*(\vec{k},M',z')j_\ell(kr)j_\ell(kr') \\
    \times Y_{\ell m}^*(\hat{\vec{k}})Y_{\ell m}(\hat{\vec{k}})\bar{n}(M,z)\bar{n}(M',z')b(M,z)b(M',z')D(z)D(z')P(\vec{k},z=0).
\end{multline}
Integrating over the direction of $\vec{k}$, and noting that in our model $\tilde{y}_{\rm 3D}$ only depends on the magnitude of $\vec{k}$ because of the spherical symmetry of our gas profile,  
\begin{multline}
    C_\ell^{yy,2h} = \frac{2}{\pi}\int adra'dr'dMdM'dkk^2\tilde{y}_\textrm{3D}(k,M,z)\tilde{y}_\textrm{3D}^*(k,M',z')j_\ell(kr)j_\ell(kr') \\
    \times \bar{n}(M,z)\bar{n}(M',z')b(M,z)b(M',z')D(z)D(z')P(k,z=0).
\end{multline}
We then use the Limber approximation \citep{1953ApJ...117..134L}, 
\begin{align}
    \int_0^\infty dk k^2 j_\ell(kr_1)j_\ell(kr_2)f(k) \approx \frac{\pi}{2r_1^2}f(\ell/r_1)\delta_D(r_1-r_2),
\end{align}
(see also \citet{2011MNRAS.418.2207T}), which gives
\begin{align}
    C_\ell^{yy,2h} &= \int a^2\frac{dr}{r^2}\int dM\tilde{y}_\textrm{3D}(k=\frac{\ell}{r},M,z)\bar{n}(M,z)b(M,z)\int dM'\tilde{y}_\textrm{3D}^*(k=\frac{\ell}{r},M',z)\bar{n}(M',z)b(M',z)P(k=\frac{\ell}{r},z) \\
    &= \int a^2\frac{dr}{r^2}\bigg\lvert \int dM\tilde{y}_\textrm{3D}(k=\frac{\ell}{r},M,z)\bar{n}(M,z)b(M,z) \bigg\rvert^2 P(k=\frac{\ell}{r},z).
\end{align}
Performing the change of variable $dr = \frac{c}{H}dz$, we obtain
\begin{align}
    C_\ell^{yy,2h} &= \int a^2\frac{c}{H}\frac{dz}{\chi^2}\bigg\lvert \int dM\tilde{y}_\textrm{3D}(k=\frac{\ell}{\chi},M,z)\bar{n}(M,z)b(M,z) \bigg\rvert^2 P(k=\frac{\ell}{\chi},z) \\
    &= \int \frac{H}{c\chi^2}a^2dz\bigg\lvert \int dM\frac{c}{H}\tilde{y}_\textrm{3D}(k=\frac{\ell}{\chi},M,z)\bar{n}(M,z)b(M,z) \bigg\rvert^2 P(k=\frac{\ell}{\chi},z) \\
    &= \int \frac{H}{c\chi^2}W_y^2(z)dz\bigg\lvert \int dMH_y(k=\frac{\ell}{\chi} \vert M,z)\bar{n}(M,z)b(M,z) \bigg\rvert^2 P(k=\frac{\ell}{\chi},z),
\end{align}
with $W_y(z)=a=(1+z)^{-1}$ and $H_y(k\vert M,z) = \frac{c}{H}\tilde{y}_\textrm{3D}(k,M,z)$.
\newline The third term in~\ref{eq:sum_3_terms}, which is the one-halo term, leads to
\begin{align}
    \langle\langle y_{\ell m}y_{\ell m}^*\rangle\rangle_3 = C_\ell^{yy,1h} = \frac{2}{\pi}\int adra'dr'dMd^3k\, \tilde{y}_\textrm{3D}(k,M,z)\tilde{y}_\textrm{3D}^*(k,M,z')j_\ell(kr)j_\ell(kr')Y_{\ell m}^*(\hat{\bold{k}})Y_{\ell m}(\hat{\bold{k}})\bar{n}(M,z).
\end{align}
Integrating over the angles of $\vec{k}$ gives 
\begin{align}
    C_\ell^{yy,1h} = \frac{2}{\pi}\int adra'dr'dMdkk^2\tilde{y}_\textrm{3D}(k,M,z)\tilde{y}_\textrm{3D}^*(k,M,z')j_\ell(kr)j_\ell(kr')\bar{n}(M,z).
\end{align}
We then use the Limber approximation
\begin{align}
    C_\ell^{yy,1h} &= \int a^2\frac{dr}{r^2}dM\tilde{y}_\textrm{3D}(k=\frac{\ell}{r},M,z)\tilde{y}_\textrm{3D}^*(k=\frac{\ell}{r},M,z)\bar{n}(M,z) \\
    &=\int a^2\frac{dr}{r^2}\int dM\bar{n}(M,z)\lvert\tilde{y}_\textrm{3D}(k=\frac{\ell}{r},M,z)\rvert^2.
\end{align}
Using again the change of variable $dr = \frac{c}{H}dz$, we obtain
\begin{align}
    C_\ell^{yy,1h} &=\int a^2\frac{c}{H}\frac{dz}{\chi^2}\int dM\bar{n}(M,z)\lvert\tilde{y}_\textrm{3D}(k=\frac{\ell}{\chi},M,z)\rvert^2 \\
    &= \int \frac{H}{c\chi^2}a^2dz\int dM \bar{n}(M,z)\bigg\lvert\frac{c}{H}\tilde{y}_\textrm{3D}(k=\frac{\ell}{\chi},M,z)\bigg\rvert^2 \\
    &= \int \frac{H}{c\chi^2}W_y^2(z)dz\int dM \bar{n}(M,z)\bigg\lvert H(k = \frac{\ell}{\chi} \vert M,z)\bigg\rvert^2,
\end{align}
with again $W_y(z)=a=(1+z)^{-1}$ and $H_y(k\vert M,z) = \frac{c}{H}\tilde{y}_\textrm{3D}(k,M,z)$.

\end{appendix}

\end{document}